\documentclass[10pt,twocolumn,letterpaper]{article}
\usepackage[margin=0.7in,columnsep=0.25in]{geometry}
\usepackage[T1]{fontenc}
\usepackage[utf8]{inputenc}
\usepackage{lmodern}
\usepackage{amsmath,amssymb,graphicx,booktabs,array,calc}
\usepackage[font=small,labelfont=bf]{caption}
\usepackage[numbers,super,sort&compress]{natbib}
\usepackage{microtype}
\usepackage[hidelinks]{hyperref}
\usepackage{url}
\date{}
\begin{document}
\twocolumn[{\begin{@twocolumnfalse}
\begin{center}
{\Large\bfseries Recursive extrema extraction tracks temporal organization through progressive data reduction\par}\vspace{9pt}
{\large Vadim Shlyonsky\par}\vspace{5pt}
Laboratoire d'Enseignement de la Physique, Faculté de Médecine,\\
Université libre de Bruxelles, Brussels, Belgium\\ \vspace{3pt}
\href{mailto:Vadim.Shlyonskiy@ulb.be}{Vadim.Shlyonskiy@ulb.be}\quad ORCID: 0000-0003-2999-6553
\end{center}
\begin{abstract}
Temporal organization is commonly characterized at a prescribed
observational scale. Here, a sequence is repeatedly reduced by retaining
only the local maxima surviving the preceding extraction. After each
step, retained maxima are re-indexed, discarding the original clock and
generating a hierarchy whose successive scales are determined by the
surviving observations. Mean and variance of gaps between local maxima
define hierarchical maxima-gap statistics (HMGS), represented as a
trajectory in a two-dimensional statistical plane. Continuous
independent and identically distributed (IID) signals generate a
distribution-independent reference trajectory, whereas the displacement
of structured processes may collapse, persist or amplify relative to it.
Crucially, collapse becomes visible only when the accessible hierarchy
reaches the characteristic scale of the organization, coupling the
required record length to the scale being tested. These fates occur
across stochastic, chaotic and event-derived systems and cannot
generally be inferred from the first level. HMGS complements analyses
based on correlations, entropy, recurrence or prescribed scales by
asking a distinct question: whether organization identified at one level
remains detectable as the observations themselves define successively
deeper levels, without fitting a generative model or prescribing an
external coarse-graining scale. In synthetic neuronal populations,
superposition creates non-IID trajectories, while source organization
persists or is attenuated according to its temporal structure. For mouse
population spike trains, median covariance-normalized separation from
matched shuffled-source controls decreased from 9.1 times intrinsic null
variability at the initial level to 1.2 times after three recursive
reductions. HMGS thereby probes which signatures of temporal
organization survive progressive data reduction.
\end{abstract}
\par\vspace{8pt}
\end{@twocolumnfalse}}]
\noindent\textbf{Temporal organization in a time series can reflect the dynamics of its source or arise when observations are processed or combined. Here, temporal organization is probed at successive levels of a hierarchy. The local maxima of the original sequence form a new sequence, whose own local maxima are selected in turn. At each level, the spacing between successive maxima is measured in the sequence being examined. The hierarchy has no preset depth. The available record length determines how many levels retain enough gaps for reliable estimation. Independent and identically distributed observations provide a reference for following how departures from randomness fade, persist or grow. In neuronal population recordings, comparison with shuffled spike trains also reveals how much organization can be attributed to the timing of events within the constituent sources.}
\par\medskip
\section*{Organization revealed by repeated reduction}

\begin{figure*}[tp]
\centering
\includegraphics[width=0.92\textwidth,height=6.9in,keepaspectratio]{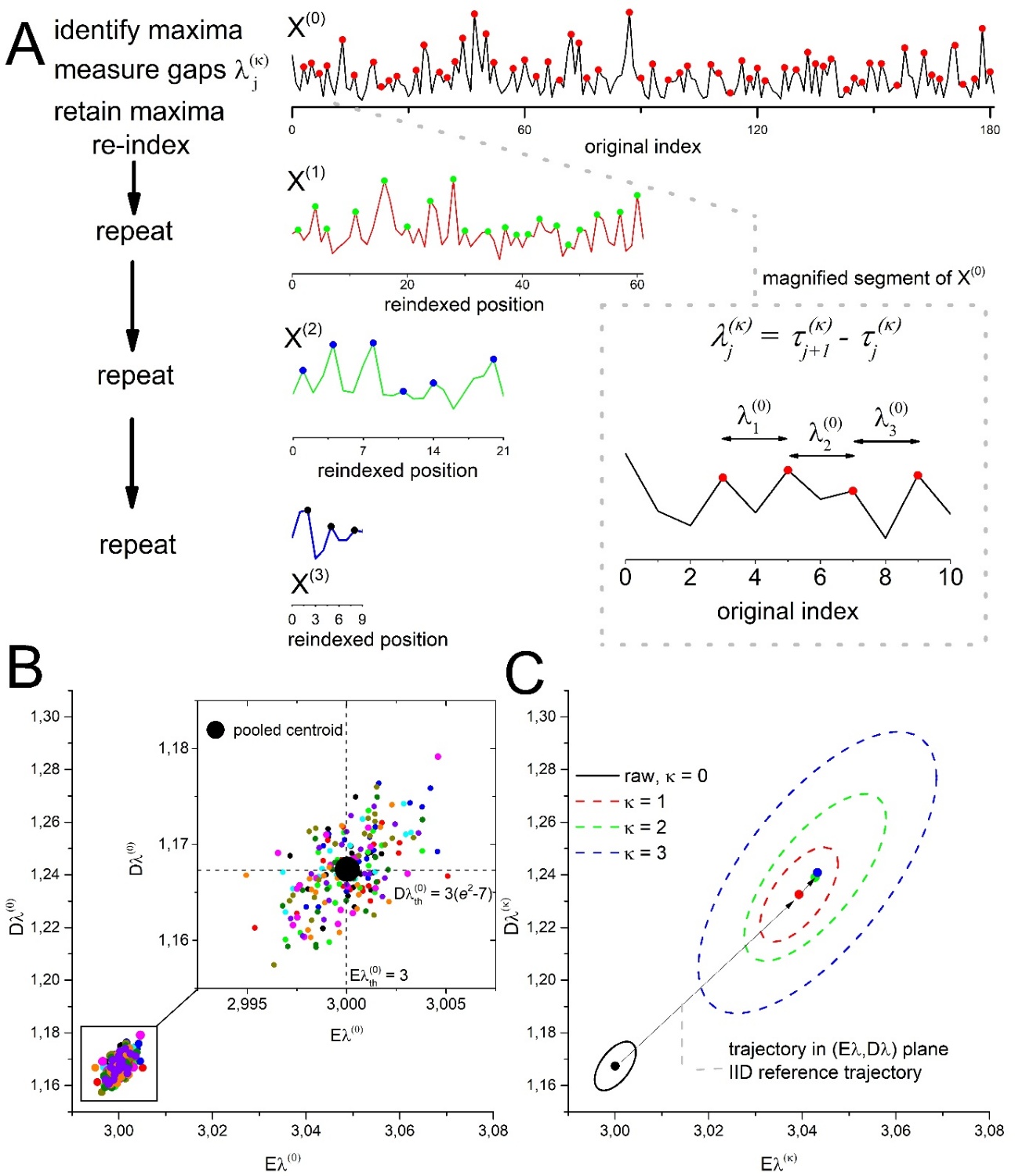}
\caption{\textbf{Recursive extrema extraction transforms temporal
organization into a trajectory. A}, At each hierarchy order \(\kappa\),
local maxima of \(X^{(\kappa)}\) are identified and the gaps between
their indices are measured as
\(\lambda_{j}^{(\kappa)} = \tau_{j + 1}^{(\kappa)} - \tau_{j}^{(\kappa)}\).
The retained maxima, preserved in their original order, form
\(X^{(\kappa + 1)}\) and are consecutively re-indexed before the
operation is repeated. Colours show how maxima selected at one level
become the sequence analysed at the next; the enlargement illustrates
gap measurement at \(\kappa = 0\). \textbf{B}, Raw gap statistics from
200 realizations representing ten continuous IID models. Colors
distinguish IID generators. The pooled centroid lies close to the exact
distribution-independent values \(E_{\lambda}^{(0)} = 3\) and
\(D_{\lambda}^{(0)} = 3\left( e^{2} - 7 \right)\). \textbf{C}, Empirical
95\% bivariate IID reference regions at hierarchy orders
\(\kappa = 0\text{–}3\). Points indicate order-specific pooled centroids
and the connecting arrow shows their trajectory, defining the IID
reference trajectory.}
\label{fig:1}
\end{figure*}

Local extrema are long-standing ordinal descriptors used in randomness
testing, event detection and nonlinear time-series
characterization\cite{ref1,ref2,ref3,ref4,ref5,ref6,ref7,ref8,ref9,ref10}. Yet temporal organization need
not provide a natural observational scale. Here, a sequence is
repeatedly reduced to its local maxima (Fig. \hyperref[fig:1]{1}A). After each extraction,
retained amplitudes are re-indexed in their original order, so
next-level gaps are measured on a scale generated by the survivors
rather than the original clock. This ordinal selection is invariant
under strictly increasing amplitude transformations but remains
sensitive to temporal ordering. At each level, gap mean and variance
define a point in the (E\ensuremath{\lambda},D\ensuremath{\lambda}) plane; successive points form a
trajectory. Recursion may continue while sufficient observations remain;
orders \ensuremath{\kappa}=0--3 are analyzed here (see Methods).

Separations between same-type turning points were examined by Rozenberg
and Ruderman and later by Rozenberg\cite{ref11,ref12}. In another
setting, Derrida and Gardner derived the distance distribution between
weak bonds in a one-dimensional spin-glass chain\cite{ref13};
Hivert et al. showed that it equals nearest-peak spacing in a
random-permutation model of ballistic surface
growth\cite{ref14}. For continuous IID input, Zubkov and
colleagues derived the distribution-independent local-maxima gap
law\cite{ref15}: the exact raw IID coordinates are E\ensuremath{\lambda}\textsuperscript{(0)}=3 and
D\ensuremath{\lambda}\textsuperscript{(0)}=3(e\textsuperscript{2}\ensuremath{-}7). Across 200 realizations spanning ten IID models,
estimates formed a narrow cloud around the analytical point despite
widely differing amplitude distributions (Fig. \hyperref[fig:1]{1}B).

\begin{figure*}[tp]
\centering
\includegraphics[width=0.92\textwidth,height=6.9in,keepaspectratio]{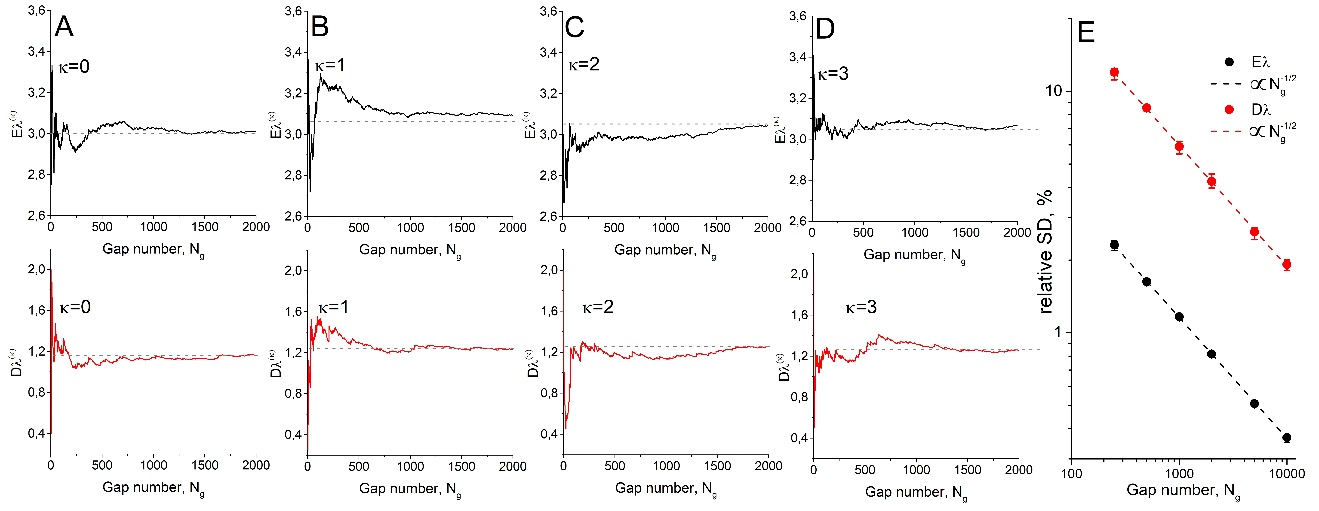}
\caption{\textbf{Finite-record behavior and precision of IID hierarchy
statistics.}\\
\textbf{A--D,} Running estimates of \(E\lambda^{(\kappa)}\) (black) and
\(D\lambda^{(\kappa)}\) (red) for one representative Gaussian IID
realization at \(\kappa = 0\text{–}3\). Dashed horizontal lines indicate
the corresponding complete-record estimates. \textbf{E,} Relative
across-realization SD of the estimated mean and variance at fixed
order-specific gap count \(N_{g}\). Symbols show means across
\(\kappa = 0\text{–}3\), asymmetric error bars span the order-wise
minimum and maximum. Dashed lines show unconstrained log-log power-law
fits:
\(Relative{SD}\left( \widehat{E} \right) = 36.5N_{g}^{- 0.500}\ \%\ \ \) and
\(Relative\ SD\left( \widehat{D} \right) = 190N_{g}^{- 0.501\ \ }\%\),
consistent with \(N_{g}^{- 1/2}\) finite-sample scaling.}
\label{fig:2}
\end{figure*}

Because the analytical coordinates are population moments, empirical
agreement is intrinsically a large-sample property. Figure \hyperref[fig:2]{2} quantifies
finite-record precision as a function of order-specific gap count:
across-realization relative SD followed 36.5
\emph{N\textsubscript{g}}\textsuperscript{-0.500}\% for
\emph{E\textsubscript{\ensuremath{\lambda}}} and 190
\emph{N\textsubscript{g}}\textsuperscript{-0.501}\% for
\emph{D\textsubscript{\ensuremath{\lambda}}} (Fig. \hyperref[fig:2]{2}E), corresponding at 2000 gaps to
approximately 0.8\% and 4.2\%, respectively. Running estimates from one
realization illustrate approach to the complete-record values but were
not used for record selection (Fig. \hyperref[fig:2]{2}A-D). The variance estimate
therefore requires substantially more gaps than the mean for the same
relative precision.
These precision estimates describe sampling dispersion of gap moments;
they do not establish freedom from finite-block bias. In fragmented
records, pooling more short blocks can improve precision while leaving
a systematic displacement from the long-series IID reference, because
only gaps bounded by two maxima within the same block are observed.
Pooled gap count and individual block length at the analyzed hierarchy
order must therefore be assessed separately (Methods, Sec. 1.3).

Recursive extraction does not, however, return repeatedly to the same
raw IID point. Maxima selection transforms an IID sequence into a weakly
dependent one: neighboring gaps share a boundary maximum, whose height
influences both adjacent separations. More precisely, the augmented
sequence containing the gap length and boundary-maximum height is
Markov, whereas the observed gap sequence alone has hidden-Markov
organization\cite{ref16}. Consequently, later hierarchy
sequences cannot be treated as fresh IID inputs to which the raw result
is simply reapplied. Distinct order-specific IID reference regions are
generated instead (Fig. \hyperref[fig:1]{1}C). Relative to the raw \(\kappa = 0\ \)point,
the \(\kappa = 1\) extraction produces a reproducible displacement,
while subsequent levels remain within a compact higher-order region. The
trajectory connecting their centroids defines the IID reference
trajectory. Structured sequences can therefore be compared with the
appropriate IID region at each order, allowing temporal organization to
be characterized not only by its initial displacement but also by how
that displacement changes under repeated reduction.

\section*{Three fates of temporal organization}

\begin{figure*}[tp]
\centering
\includegraphics[width=0.92\textwidth,height=6.2in,keepaspectratio]{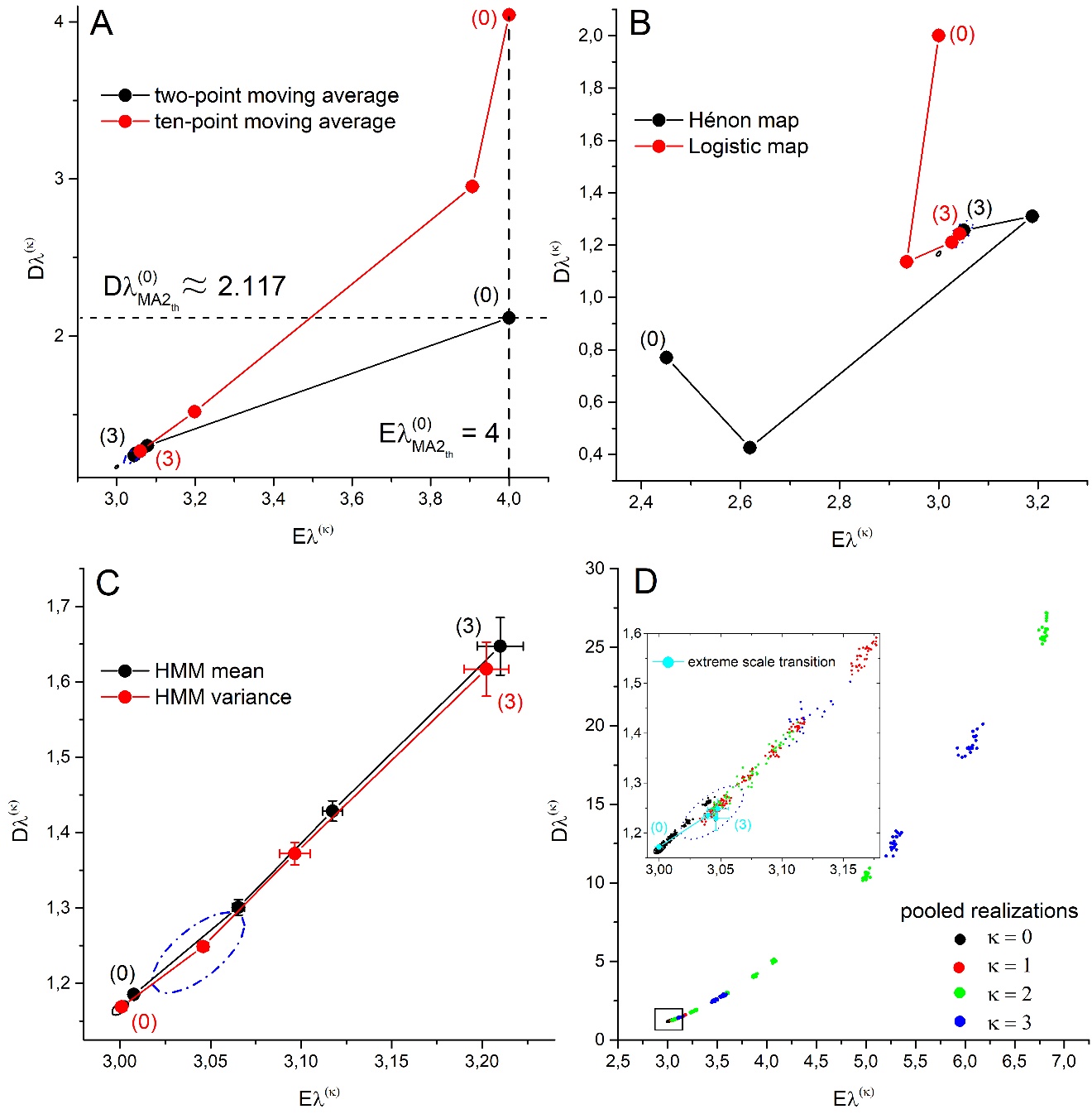}
\caption{\textbf{Temporal organization collapses or amplifies under
recursive extrema extraction. A}, Hierarchy trajectories of two- and
ten-point moving averages applied to IID input. Dashed lines indicate
the exact raw mean E\ensuremath{\lambda}\textsuperscript{(0)} = 4 and analytical variance D\ensuremath{\lambda}\textsuperscript{(0)} \ensuremath{\approx} 2.117 for
the two-point average. The two-point trajectory collapses rapidly toward
the higher-order IID reference region, whereas the ten-point trajectory
remains strongly displaced at intermediate orders before approaching the
reference at \ensuremath{\kappa} = 3, demonstrating that collapse is delayed as the
averaging window increases. \textbf{B}, Hierarchy trajectories of the
logistic and Hénon maps. Both collapse toward the higher-order reference
by different routes: the logistic map reaches it at \ensuremath{\kappa} = 2, whereas the
Hénon map reaches it at \ensuremath{\kappa} = 3 after a broad intermediate excursion.
\textbf{C}, Mean-switching and variance-switching hidden Markov Gaussian
models with P\textsubscript{stay} = 0.995. Both trajectories move
progressively away from the reference regions, demonstrating
amplification of persistent hidden-state organization. \textbf{D},
Pooled hierarchy statistics for Gaussian and exponential IID signals
contaminated by monotonic trends of increasing strength. Successive
orders generate large outward excursions. The inset enlarges the
low-deviation region and shows an extreme exponential scale transition,
which remains close to the order-matched IID reference despite its large
amplitude change. In A--C, points show means across 20 independent
realizations; error bars indicate SD and, where not visible, are smaller
than the symbols. In D, points represent individual realizations. Solid
black and dash--dotted blue ellipses indicate the empirical IID
reference regions at \ensuremath{\kappa} = 0 and \ensuremath{\kappa} = 3, respectively.}
\label{fig:3}
\end{figure*}

\begin{figure*}[tp]
\centering
\includegraphics[width=0.92\textwidth,height=6.2in,keepaspectratio]{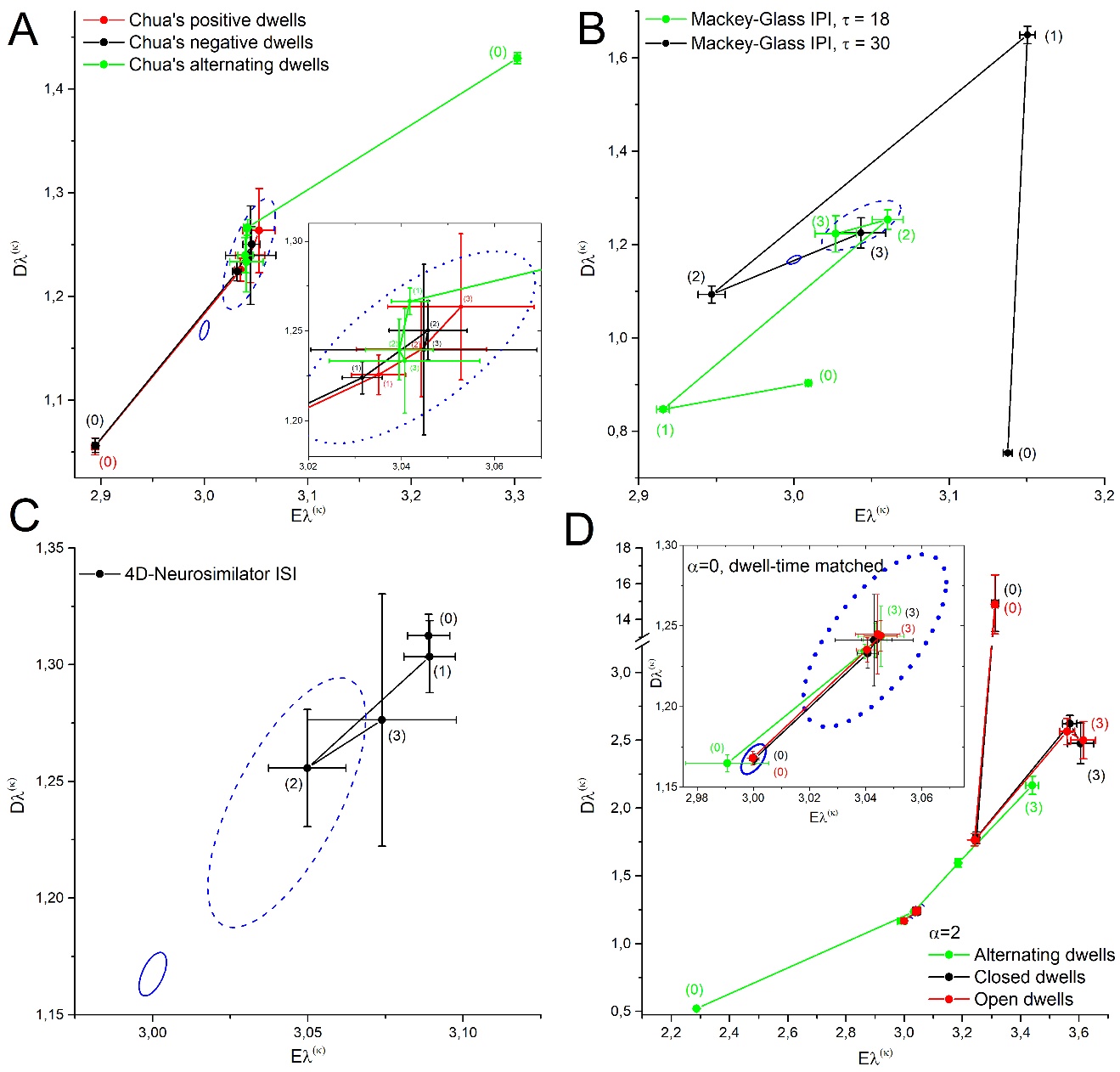}
\caption{\textbf{Event intervals reveal collapse and amplification of
temporal organization. A}, Positive, negative and alternating dwell
sequences obtained from the chaotic Chua circuit; the inset enlarges the
higher-order region. \textbf{B}, Inter-peak interval (IPI) sequences
from Mackey--Glass oscillations with delays \ensuremath{\tau}=18 and 30. \textbf{C},
Inter-spike interval (ISI) sequences from a nonlinear four-dimensional
electronic spiking neuron. In A--C, distinct low-order excursions are
progressively reduced and the trajectories reach or approach the
higher-order IID region. \textbf{D}, Open, closed and alternating dwell
sequences from the strong membrane-memory model (\ensuremath{\alpha}=2, \ensuremath{\rho}=0.995 and
\ensuremath{\sigma}=0.08); the ordinate is interrupted between the low- and high-variance
regions. The inset shows memory-free \ensuremath{\alpha}=0 controls whose state-specific
mean dwell durations were matched to the \ensuremath{\alpha}=2 condition
(\textbar \ensuremath{\Delta}\ensuremath{\tau}\textsubscript{o-C}\textbar=10.77\ensuremath{\pm}9.91 sampling intervals).
Their confinement to the IID regions separates the effect of temporal
modulation from that of mean open--closed dwell asymmetry. Points show
means across independent realizations (n=20 in A,B,D; n=10 in C); error
bars indicate SD and, where not visible, are smaller than the symbols.
Solid and dotted blue ellipses indicate the empirical IID reference
regions at \ensuremath{\kappa}=0 and \ensuremath{\kappa}=3, respectively.}
\label{fig:4}
\end{figure*}

Position relative to the order-matched IID reference is conveyed by the
hierarchy trajectory rather than any coordinate alone. The analysis asks
when displacement appears and whether it collapses, persists or
amplifies over observed orders. These are empirical responses of the
regimes examined, not permanent classes of systems. Figure \hyperref[fig:3]{3} tests them
before extension to event-derived intervals in Fig. \hyperref[fig:4]{4}.

Moving-average filtering provides an analytically anchored example of
scale-dependent collapse, echoing smoothing's capacity to create
apparent structure from random input\cite{ref17} (Fig. \hyperref[fig:3]{3}A). A
two-point average shifts raw statistics to the exact mean
\(E\lambda^{(0)} = 4\) and variance
\(D\lambda^{(0)} \approx 2.117\)\cite{ref15}, but returns to the
higher-order IID region by \(\kappa = 3\). A ten-point average makes a
larger, longer excursion before approaching the reference at
\(\kappa = 3\). Larger windows delay this collapse: MA100 begins only at
\(\kappa = 3\), whereas MA1000 does not collapse through \(\kappa = 3\)
(Fig. \hyperref[fig:S1]{S1}a,b). Thus, low-order displacement does not determine fate;
attenuation appears only when hierarchy depth becomes sufficient
relative to the imposed scale. Keras-style overlapping-window
transformations reproduced this dependence (Fig. \hyperref[fig:S1]{S1}c--f): window means
gave moving-average-like trajectories, maxima a distinct manifold with
larger excursions, and the last-sample control remained IID-like. Mean
and maximum displacements were strongest at stride 1 and disappeared
when stride equaled the window length, showing that overlap created
organization while the reduction rule determined trajectory geometry.

Less direct collapse routes are followed by deterministic chaotic maps
(Fig. \hyperref[fig:3]{3}B). The logistic-map trajectory begins with strongly inflated
variance, moves below the IID region at \ensuremath{\kappa}=1 and reaches the higher-order
reference by \ensuremath{\kappa}=2. The Hénon map begins below and to the left of the raw
reference, undergoes a broad intermediate excursion and reaches the
higher-order region only at \ensuremath{\kappa}=3. Markedly different starting coordinates
and paths can therefore lead to the same projected reference region. The
depth at which collapse is reached, and the route taken before it is
reached, contain information absent from either the starting or final
coordinate alone. Importantly, arrival at the higher-order IID region
indicates loss, under recursive reduction, of the organization detected
by these gap statistics; it does not imply that the generating dynamics
are IID or that other parameter regimes of the same system must follow
the same hierarchy trajectory.

The opposite limiting response is produced by sufficiently persistent
hidden-state organization (Fig. \hyperref[fig:3]{3}C). In mean-switching and
variance-switching hidden Markov Gaussian models with
P\textsubscript{stay}=0.995, the raw statistics remain close to the IID
region, but successive hierarchy orders move progressively outward.
Temporal organization that is scarcely expressed at \ensuremath{\kappa}=0 can therefore
become visible only after recursive reduction. Amplification depends on
state persistence: less persistent hidden regimes produce transient
departures that subsequently collapse toward the reference (Fig.
\hyperref[fig:S2]{S2}a--c). Additional Markov and nonlinear dependencies generated distinct
order-dependent trajectories (Fig. \hyperref[fig:S2]{S2}d--f).

Monotonic trends of increasing strength added to Gaussian and
exponential IID signals also produce large outward trajectories (Fig.
\hyperref[fig:3]{3}D; Fig. \hyperref[fig:S3]{S3}). By contrast, an extreme exponential scale transition
remains near the order-matched references despite its large amplitude
change (Fig. \hyperref[fig:3]{3}D, inset): amplitude change alone need not cause
amplification, whereas the trend persistently biases local ordering.

The observed response depends on the relation between the characteristic
scale of the organization and the effective depth reached by the gap
hierarchy. When that scale is reached, displacement may progressively
diminish; when it is not, the trajectory may remain displaced throughout
the available orders, as for MA1000. Sufficiently persistent latent
states or directional ordering may instead become increasingly prominent
among the surviving maxima. A reproducible displacement that survives
without systematic outward growth defines the intermediate case of
persistence. Coefficient-dependent autoregressive and moving-average
manifolds illustrate this response (Fig. \hyperref[fig:S4]{S4}). These terms describe
behavior only over the observed hierarchy. Similar trajectories need not
identify a common mechanism: hidden-state memory and directional
non-stationarity both amplify in Fig. \hyperref[fig:3]{3}C,D, making matched controls
essential for physical interpretation. The same scale--depth dependence
was observed for sinusoidal contamination: the trajectory depended
jointly on sinusoid period and the sinusoid-to-source standard-deviation
ratio (Fig. \hyperref[fig:S5]{S5}). Fitting and subtracting the sinusoid restored the
uncontaminated IID reference trajectory and left IID controls unchanged.
The scope of this validation was limited to sinusoidal contamination of
IID signals; linear detrending and decontamination of temporally
organized signals were not examined.

\section*{From sampled signals to event intervals}

Many physical and biological records are more naturally represented by
events than by continuously sampled amplitudes. Switching circuits yield
dwell times, oscillators inter-peak intervals and neurons inter-spike
intervals. Thresholding, peak detection or state segmentation therefore
precedes hierarchy analysis and produces an ordered interval sequence.
Recursive extraction is then applied to that sequence, not to the
original waveform: for a spike train, \ensuremath{\kappa}=0 maxima are inter-spike
intervals longer than the preceding interval and at least as long as the
following interval, and gaps between them are measured as differences
between their indices in the interval sequence. Event definition and
hierarchy analysis are thus separate operations.

Different interval representations of the same dynamics need not begin
at the same coordinate. In the Chua circuit, statistics for positive and
negative dwell sequences follow nearly coincident trajectories, whereas
the complete alternating sequence begins in a distinct region because
the succession of both states is retained (Fig. \hyperref[fig:4]{4}A). Nevertheless, all
three approach the higher-order IID region. Mackey--Glass inter-peak
intervals at \ensuremath{\tau}=18 and \ensuremath{\tau}=30 follow markedly different, non-monotonic
routes before reaching the same region by \ensuremath{\kappa}=3 (Fig. \hyperref[fig:4]{4}B). The nonlinear
four-dimensional electronic spiking neuron showed a comparable but
slower contraction (Fig. \hyperref[fig:4]{4}C). Across ten SPICE simulations at different
input-current operating points within the thalamic-type
irregular-spiking regime\textsuperscript{18}, displaced raw ISI statistics moved toward
the higher-order reference region. Variability increased at \ensuremath{\kappa}=3, but the
trajectory direction was preserved. Event-derived representations
therefore retain system-specific organization at low orders, yet that
organization can share the same fate of collapse.

Persistent memory, reported in several ion-channel dwell sequences but
still debated in relation to Markov and non-Markov
descriptions\cite{ref19,ref20}, produces a different response. In
the membrane-memory model, a slowly varying hidden environment modulates
the local open- and closed-dwell scales in opposite directions. Under
strong coupling (\ensuremath{\alpha}=2), open, closed and alternating sequences move away
from the reference after passing through its vicinity at \ensuremath{\kappa}=1, so memory
becomes increasingly visible among the surviving events (Fig. \hyperref[fig:4]{4}D). By
contrast, \ensuremath{\alpha}=0 controls with state-specific mean dwell times matched to
the strong-memory condition remain within the order-matched IID regions
(Fig. \hyperref[fig:4]{4}D, inset; Fig. \hyperref[fig:S6]{S6}). The amplification is therefore not explained
by the modest difference between mean open and closed dwell times. More
generally, the starting coordinate does not determine whether
organization in an event sequence will collapse or be amplified by
recursion.

Because events are inferred from measurements, apparent interval
organization can be created or removed during extraction. Genuine short
events can be merged when the sampling interval is too coarse or
low-pass filtering is too aggressive, whereas false threshold crossings
can be introduced at low signal-to-noise ratios\cite{ref21,ref22,ref23,ref24}.
Under the tested conditions, amplitude quantization had negligible
effects while the deepest-order adjacent-tie fraction remained about 3\%
or less; canonicalizing Mackey--Glass intervals to the nominal 0.1-unit
integration grid likewise left their hierarchy trajectories effectively
unchanged (see Methods). Even without extraction errors, deterministic
alternation of independently drawn dwell times from unequal open and
closed distributions can displace the raw statistics from the IID
reference; under the conditions tested, however, these deviations
collapse toward the higher-order reference region. Independent
asymmetric dwell distributions supplied memory-free controls (Fig. \hyperref[fig:S7]{S7}),
whereas ground-truth, noisy and filtered trace reconstructions,
event-count and dwell-duration distortions, and reciprocal baseline
asymmetry were evaluated separately (Figs. \hyperref[fig:S8]{S8}, \hyperref[fig:S9]{S9}, \hyperref[fig:S10]{S10} and \hyperref[fig:S11]{S11}). When the shortest
relevant events are adequately resolved, hierarchy analysis can then be
applied to population spike trains, but interpretation requires
source-matched superposition nulls.

\textbf{Superposition creates a non-IID background and reshapes source
organization.}

When event sequences are combined, a population spike train is formed by
temporal interleaving rather than linear addition. Superposition of
independent renewal spike trains need not yield a renewal or Poisson
process and can introduce serial interval correlations absent from the
sources\cite{ref25,ref26}. A displaced population coordinate
therefore cannot by itself identify inherited organization. Matched
nulls must reproduce the constituent interval distributions and the
complete superposition procedure while removing only source-level serial
order.

\begin{figure*}[tp]
\centering
\includegraphics[width=0.92\textwidth,height=6.2in,keepaspectratio]{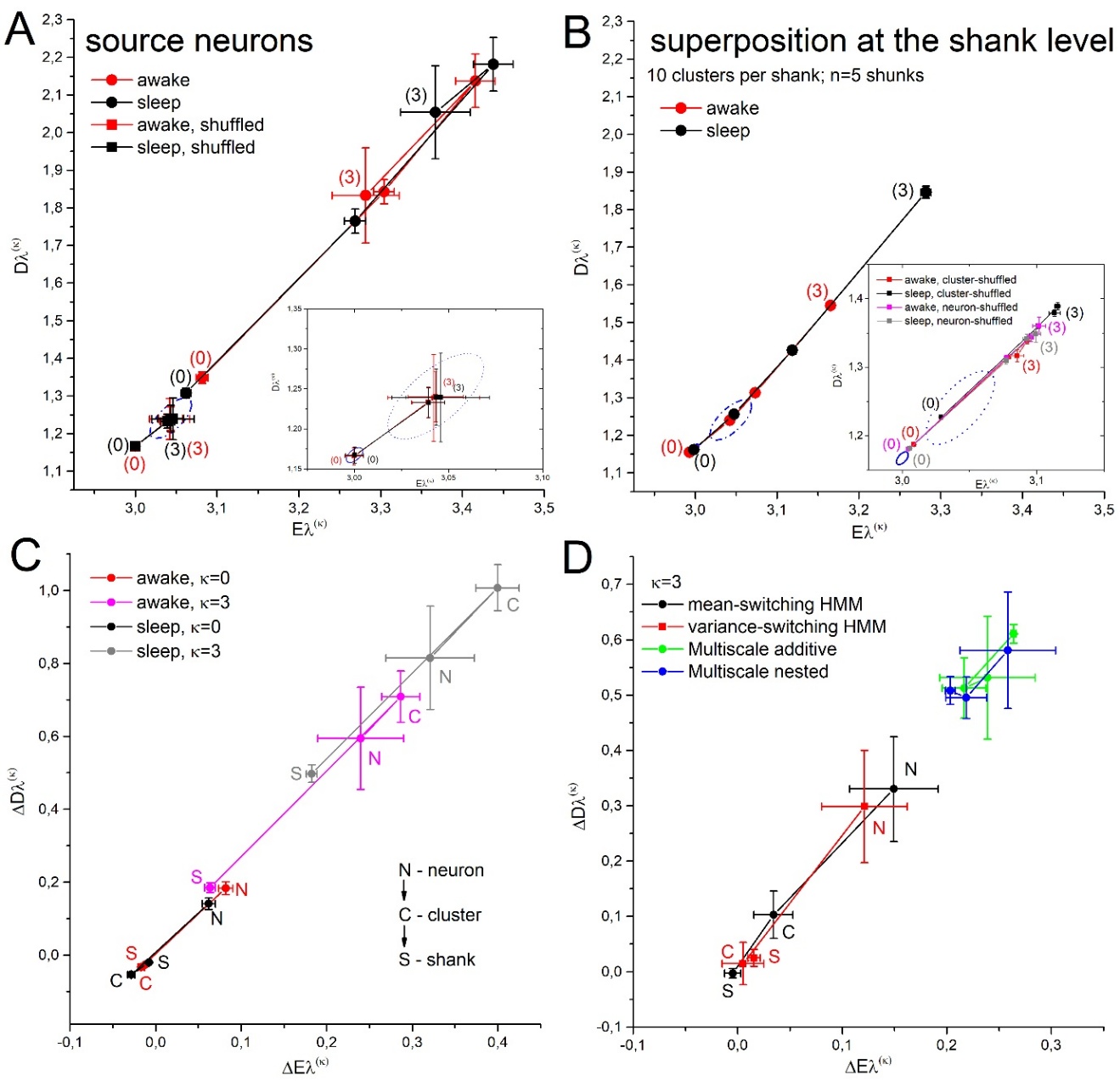}
\caption{\textbf{Superposition creates a non-IID background while
selectively preserving source organization. A}, Hierarchy trajectories
of inter-spike intervals from awake-like and sleep-like AR-modulated
source neurons and their within-neuron shuffled controls. \textbf{B},
Organized population sequences formed by pooling five neurons per
cluster and ten clusters per shank. The inset shows matched shank nulls
constructed either by shuffling each source-neuron interval sequence
before pooling or by shuffling the ten organized cluster interval
sequences before shank formation. \textbf{C}, Paired
organized-minus-neuron-shuffled residuals at the neuron (N), cluster (C)
and shank (S) levels, shown at \ensuremath{\kappa}=0 and \ensuremath{\kappa}=3; the origin denotes agreement
with the matched null. \textbf{D}, Residuals at \ensuremath{\kappa}=3 for mean-switching
and variance-switching hidden-Markov populations and for additive and
nested multiscale populations. Lines connect N\ensuremath{\rightarrow}C\ensuremath{\rightarrow}S. Hidden-Markov
residuals approach the origin with aggregation, whereas multiscale
residuals remain displaced. Points in A and B show means across 250
neurons and five shanks, respectively. Points in C and D show mean
paired residuals, with n=250 neuron, n=50 cluster and n=5 shank pairs;
error bars indicate SD. Solid and dotted blue ellipses in A and B
indicate the empirical IID reference regions at \ensuremath{\kappa}=0 and \ensuremath{\kappa}=3,
respectively; labels identify the displayed hierarchy orders.}
\label{fig:5}
\end{figure*}

Awake-like and sleep-like AR-modulated source neurons showed outward
hierarchy trajectories, whereas their within-neuron shuffled controls
followed the order-matched IID reference trajectory (Fig. \hyperref[fig:5]{5}A). When five
neurons were pooled into each cluster and ten clusters into a shank, the
organized shanks remained strongly displaced (Fig. \hyperref[fig:5]{5}B). The
neuron-shuffled and cluster-shuffled shank nulls were also displaced
after superposition, demonstrating that pooling itself generated non-IID
trajectories. Their close agreement further indicates that the two
tested null constructions produced similar shank references. Complete
organized and source-shuffled trajectories at the neuron, cluster and
shank levels are shown in Fig. \hyperref[fig:S12]{S12}.

The component inherited from source ordering was quantified by the
paired residual for each organized coordinate within the same unit,
realization and hierarchy order. At \ensuremath{\kappa}=0, these residuals were small
relative to those at \ensuremath{\kappa}=3 and approached the origin after shank formation
(Fig. \hyperref[fig:5]{5}C). At \ensuremath{\kappa}=3, substantial residual organization remained at the
neuron and cluster levels, particularly in the sleep-like condition, but
was strongly attenuated after pooling into shanks. Thus, much of the
large absolute shank excursion in Fig. \hyperref[fig:5]{5}B is reproduced by superposition
alone, while a smaller state-dependent component is inherited from
source temporal ordering.

Survival of source organization under aggregation depended on its
temporal structure (Fig. \hyperref[fig:5]{5}D). Aggregation of heterogeneous short-memory
processes can generate longer-lived dependence\cite{ref27,ref28},
although those process-level constructions differ from the
discrete-event interleaving examined here. Neural activity itself spans
multiple temporal scales\cite{ref29}. Accordingly, multiscale
modulation remained strongly displaced across neurons, clusters and
shanks, whereas mean- and variance-switching hidden-Markov residuals
diminished after cluster formation and approached the
superposed-shuffled controls at the shank level. Population aggregation
can therefore preserve or suppress source organization while generating
its own background trajectory. Corresponding complete multiscale and
hidden-Markov trajectories are shown in Fig. \hyperref[fig:S13]{S13} and Fig. \hyperref[fig:S14]{S14}, and their
order-resolved residuals are shown in Fig. \hyperref[fig:S15]{S15}. The multiscale-shank
classification was unchanged by continuous-time or 20-kHz pooling (see
Methods), consistent with the local effect of sparse merger elimination.

\textbf{Excess organization in the analyzed mouse recordings is
concentrated at low hierarchy orders}

Publicly available extracellular recordings from freely moving mice
(Buzsáki Laboratory th-1 dataset) were then tested for organization
beyond that created by spike-train superposition alone. For each
electrode shank recorded during slow-wave sleep in freely moving mice,
spikes from retained sorted clusters were pooled into a population
sequence. In the matched null, each cluster's inter-spike intervals were
shuffled independently within the same sleep intervals before
reconstruction and superposition. Source-specific interval
distributions, population composition and superposition-induced
structure were therefore preserved, while original within-source
ordering was removed.

\begin{figure*}[tp]
\centering
\includegraphics[width=0.92\textwidth,height=6.9in,keepaspectratio]{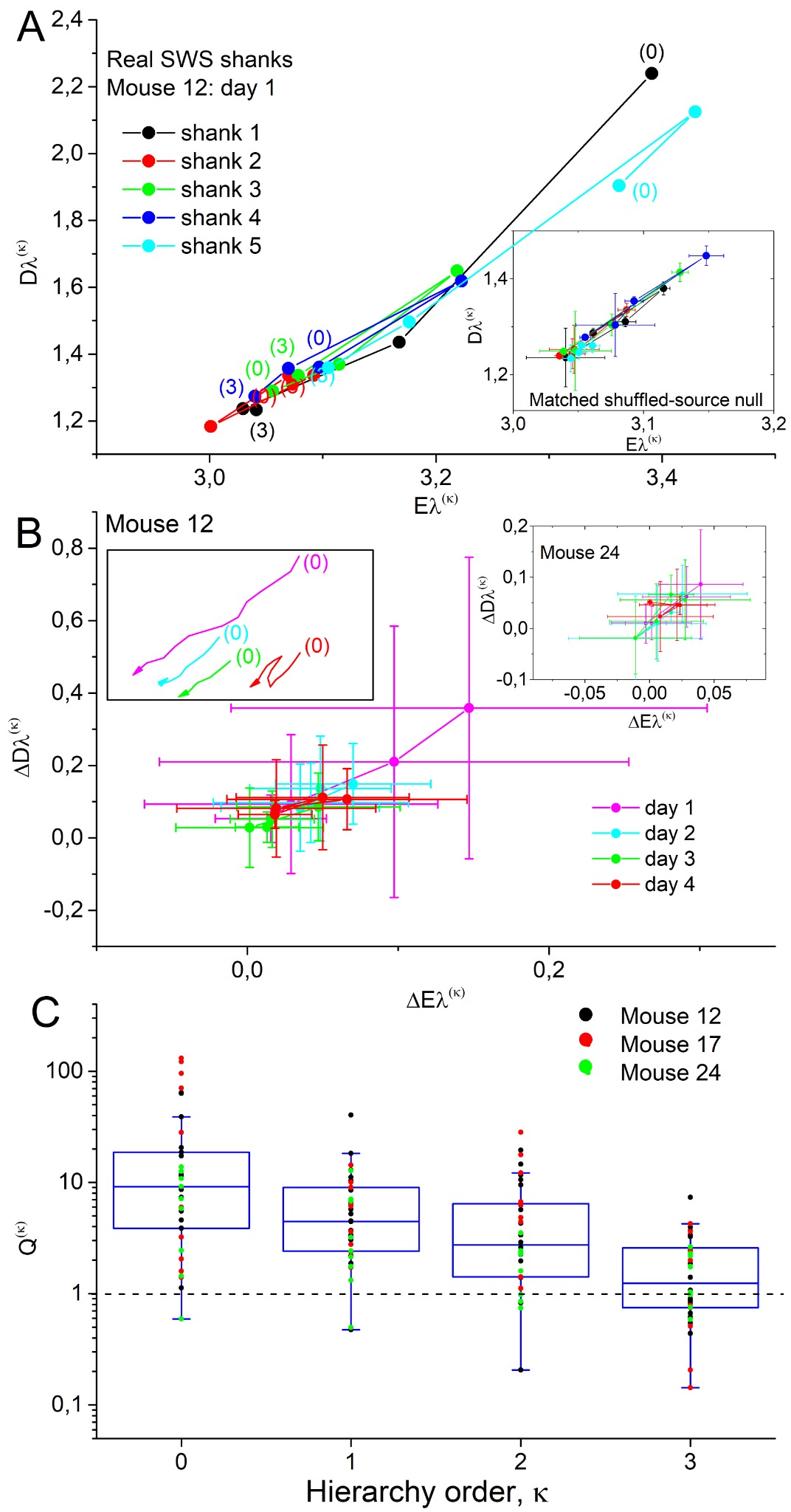}
\caption{\textbf{Excess organization in mouse population spike trains
is concentrated at low hierarchy orders.} \textbf{A}, Hierarchy
trajectories from five shanks during slow-wave sleep in Mouse 12, day 1.
Population spike trains were formed by pooling retained clusters within
each shank; labels mark \ensuremath{\kappa}=0 and \ensuremath{\kappa}=3. The inset shows matched
shuffled-source nulls constructed by independently shuffling each
cluster's inter-spike intervals within the same sleep intervals before
reconstruction and superposition. Null points show means over five
shufflings; error bars indicate SD. \textbf{B}, Paired real-minus-null
residual trajectories averaged across eligible shanks for individual
Mouse 12 recording days (colors). Error bars indicate SD across shanks.
The schematic inset separates overlapping trajectories, labels their \ensuremath{\kappa}=0
origins and uses arrowheads to indicate progression toward \ensuremath{\kappa}=3; the data
inset shows Mouse 24 on expanded axes. \textbf{C}, Normalized
displacement Q for the 38 day \ensuremath{\times} shank records with at least
2,000 pooled within-bout gaps at \ensuremath{\kappa}=3 in the recording and in each
of its five matched shuffles. Q is the Mahalanobis
real-to-null distance divided by the median leave-one-out Mahalanobis
distance among matched nulls, using the order-specific covariance of
pooled within-shank null deviations. Points show individual records,
colors denote animals, and boxes show medians and interquartile ranges.
The dashed line marks Q=1, at which real--null separation equals typical
intrinsic null variability.}
\label{fig:6}
\end{figure*}

In the representative Mouse 12 recording, real shanks followed broad,
heterogeneous trajectories, whereas matched shuffled-source populations
occupied a more restricted region (Fig. \hyperref[fig:6]{6}A). Absolute displacement from
the IID ellipses is therefore not the decisive comparison: each real
coordinate was paired with the centroid of its own matched-null
realizations. The resulting real-minus-null residual places the
expectation from source marginals plus superposition at the origin.
Day-wise residual trajectories for Mouse 12 were generally positive in
both coordinates at low orders but contracted toward the origin, with
marked variation among days and shanks (Fig. \hyperref[fig:6]{6}B). The schematic inset
shows the same hierarchy direction without error bars; an expanded Mouse
24 example illustrates the smaller residual scale in that animal.
Across-day residual trajectories for Mouse 17 and Mouse 24 are shown in
Fig. \hyperref[fig:S16]{S16}. These absolute residuals describe the geometry and
heterogeneity of contraction but do not account for differences in the
intrinsic precision of the matched nulls.

Of the 64 day \ensuremath{\times} shank records, 38 yielded at least 2,000
within-bout gaps at \ensuremath{\kappa}=3 in both the recording and each matched
shuffle. For these 38 records, real--null separation was measured at
each order by a Mahalanobis distance,
using the covariance of pooled within-shank null deviations to account
for the elliptical, correlated variability of the matched controls. The
distance was normalized by the intrinsic leave-one-out Mahalanobis
variability among the corresponding matched nulls. The median ratio Q
decreased from 9.11 at \ensuremath{\kappa}=0 to 4.46, 2.75 and 1.24 at \ensuremath{\kappa}=1, 2 and 3,
respectively (Fig. \hyperref[fig:6]{6}C). Real population activity was therefore strongly
separated from its matched superposition null at low order, whereas by
\ensuremath{\kappa}=3 the median separation approached intrinsic null variability.
Individual records nevertheless remained heterogeneous, and some
retained higher-order excess organization.

The predominantly contracting residuals show that real--null separation
is greatest at low orders. Contraction is not an inevitable consequence
of aggregation or reduction: shared multiscale modulation survived both
in synthetic populations (Fig. \hyperref[fig:5]{5}D). It therefore indicates that most
ordering-dependent organization in these recordings is less persistent
across the observed hierarchy. The endpoint is not IID but the
source-matched superposition null, which removes within-source ordering
yet cannot distinguish intrinsic memory from coordination driven by
shared input.

\section*{Why following the trajectory matters}

Recursive extrema extraction projects each order onto two analytically
anchored and finite-record-calibrated gap moments. This deliberately
low-dimensional representation neither specifies the full gap
distribution nor uniquely identifies mechanism. Interpretation therefore
depends on scale matching: short-span organization may collapse at
shallow orders, whereas longer-scale structure can remain displaced
until its characteristic scale is reached. Non-collapse establishes
persistence only across the accessible hierarchy, not scale-free or
indefinite persistence. Each extraction retains about one third
observations, so every additional order requires roughly threefold more
initial data for the same terminal gap count; structured sequences
require empirical assessment of this ratio. Temporal scale, hierarchy
depth and record length are inseparable design variables whose
appropriate combination cannot generally be specified \emph{a priori}
and must be established through empirical iteration.

Recursive removal of observations resembles coarse-graining but is not a
formal renormalization scheme: no block size or external scale factor is
imposed, the mean--variance projection is not closed, and neither a
fixed point nor a universality class is established. Re-indexing instead
creates an endogenous hierarchy whose next scale is set by the surviving
extrema. The transformation is therefore portable across sampled
signals, dwell sequences, inter-event intervals and population spike
trains.

Recursive gap analysis complements established time-series approaches
(Table~\ref{tab:comparison}). Without ties, raw maxima are \(+ - \ \)motifs in the
increment-sign sequence used by entropy of difference, and gaps are
motif return times.\cite{ref30} Recursion additionally uses
retained-maxima amplitudes, absent from binary signs, and re-indexes
survivors. Because extrema selection and recursion are nonlinear
functions of temporal order, it can expose multi-point ordinal
dependence missed by pairwise autocorrelation, including when linear
autocorrelation is zero (Fig. \hyperref[fig:S2]{S2}f). It asks whether organization remains
visible after repeated reduction to a sequence's own extrema. It can
also compare regimes: raw trajectory differences describe total change,
whereas differences between regime-specific null residuals isolate
changes in temporal ordering (see Methods for detailed explanations).
These phenomenological trajectories are not unique fingerprints of
mechanism.
\begin{table*}[tp]
\caption{\textbf{ Complementary views of temporal organization.} Each
approach resolves a different structural question; the comparison is
conceptual rather than a performance ranking.}
\label{tab:comparison}
\footnotesize
\renewcommand{\arraystretch}{1.22}
\begin{tabular}{@{}
  >{\raggedright\arraybackslash}p{(\textwidth - 6\tabcolsep) * \real{0.2185}}
  >{\raggedright\arraybackslash}p{(\textwidth - 6\tabcolsep) * \real{0.3088}}
  >{\raggedright\arraybackslash}p{(\textwidth - 6\tabcolsep) * \real{0.2028}}
  >{\raggedright\arraybackslash}p{(\textwidth - 6\tabcolsep) * \real{0.2699}}@{}}
\toprule\noalign{}
\begin{minipage}[b]{\linewidth}\raggedright
\textbf{Method family}
\end{minipage} & \begin{minipage}[b]{\linewidth}\raggedright
\textbf{Structure emphasized}
\end{minipage} & \begin{minipage}[b]{\linewidth}\raggedright
\textbf{Typical representation}
\end{minipage} & \begin{minipage}[b]{\linewidth}\raggedright
\textbf{Complementary focus of recursive extrema extraction}
\end{minipage} \\
\midrule\noalign{}

Linear correlation and spectral methods\cite{ref32} & Pairwise
linear dependence across lags or frequencies & Autocorrelation and power
spectrum & Probes multi-point ordinal structure that can remain when
linear correlation vanishes \\
Detrended fluctuation analysis\cite{ref33} & Scale-dependent
fluctuation correlations after detrending & Fluctuation function and
scaling exponent & Uses a data-generated hierarchy rather than
prescribed windows or a scaling exponent \\
Lyapunov analysis\cite{ref34} & Exponential separation of nearby
trajectories & Reconstructed state space and Lyapunov spectrum & Applies
to stochastic and event sequences without assuming deterministic
chaos \\
Entropy of difference and ordinal and amplitude-sensitive pattern
methods\cite{ref9,ref30,ref35,ref36,ref37,ref38,ref39,ref40} & Fixed-length ordinal motifs and
transitions, optionally amplitude-weighted & pattern probabilities,
transitions and entropy measures. & Recurses on selected extrema,
re-indexing survivors and following extrema-gap moments across
data-generated levels \\
Recurrence analysis\cite{ref41} & Recurrent proximity of states
or trajectories & Recurrence plots and recurrence quantification &
Tracks the fate of extrema spacing rather than recurrent proximity of
states \\
Multiscale entropy\cite{ref42,ref43} & Irregularity after
coarse-graining at prescribed scales & Entropy as a function of scale &
Generates hierarchy scales endogenously from the extrema that survive \\
Higher-order and iterative approaches\cite{ref44,ref45} &
Multivariate higher-order dependencies or distinct fluctuation layers &
Higher-order interactions, layers or components & Tests one ordered
sequence as information is progressively removed \\
\bottomrule
\end{tabular}

\end{table*}

Finite records limit which organizational scales can be tested, a
general constraint on inference from real-world
series\cite{ref31}. Gap-count targets must therefore be chosen
relative to the effect size and deepest interpreted order, with
uncertainty and order-specific gap counts reported for each application,
as explained in Methods.
For fragmented records, these requirements also apply to the lengths of
the individual contributing blocks: a large pooled sample can yield
precise estimates of boundary-biased moments. Interpretation should
therefore be anchored to a fragmentation-matched null when the
long-series reference is not demonstrably applicable.

Together, these constraints define the hierarchy's interpretable domain.
Within it, recursive reduction reveals whether displacement from an
appropriate reference contracts, persists or amplifies over the observed
orders. Temporal organization is thus characterized by both its initial
displacement and its observed fate. In sufficiently long extracellular
recordings stratified by behavior, the same comparison could test
whether learning shifts population organization towards coordination
persisting deeper in the hierarchy, without presuming a specific
mechanism.

\section*{Materials and Methods}

Python codes used for data generation and data analysis in the study can
be found on the website
\url{https://gitlab.ulb.be/vshlyons/hmgs-hierarchical-maxima-gap-statistics}

\subsection*{1.1 Hierarchical local-maxima gap analysis}

For a scalar sequence
\(X^{(0)} = \left\{ x_{i} \right\}_{i = 1}^{N_{0}},\) an interior sample
was classified as a local maximum when

\[x_{i - 1}\  < \ x_{i}\  \geq \ x_{i + 1}\]

This asymmetric rule retained the first point of a maximum plateau;
sequence endpoints were excluded. If \(\tau_{j}^{(\kappa)}\) denotes the
ordered index of maximum \emph{j} at hierarchy order \ensuremath{\kappa}, the local-maxima
gaps were

\[\lambda_{j}^{(\kappa)}\  = \ \tau_{j + 1}^{(\kappa)}\  - \ \tau_{j}^{(\kappa)},\ \ \ \ \ j\  = \ 1,\ \ldots,\ M_{\kappa}\  - \ 1.\]

The amplitudes at these maxima formed the next sequence,

\[X^{\kappa + 1}\  = \ \{ X_{\tau_{1}^{(\kappa)}}^{\kappa},\ X_{\tau_{2}^{(\kappa)}}^{\kappa},\ \ldots,\ X_{\tau_{M_{\kappa}}^{(\kappa)}}^{\kappa}\}.\]

which was re-indexed consecutively before maxima and gaps were extracted
again. Recursion continued through \ensuremath{\kappa} = 3. Thus,
\(\lambda_{j}^{(\kappa)}\) measures separation in numbers of elements at
level \ensuremath{\kappa}, not elapsed time or separation on the original axis. For an
event-based data series, \ensuremath{\kappa} = 0 maxima are inter-event intervals longer
than the preceding interval and at least as long as the following
interval, and gaps between them are measured as differences between
their indices in the interval sequence.

For n\textsubscript{\ensuremath{\kappa}} = M\textsubscript{\ensuremath{\kappa}} \ensuremath{-} 1 gaps, the mean and
unbiased sample variance were

\[{E\lambda}^{(\kappa)}\  = \ \frac{1}{n_{\kappa}}\ \sum_{j = 1}^{n_{\kappa}}\lambda_{j}^{(\kappa)},\]

\[{D\lambda}^{(\kappa)}\  = \ \frac{1}{n_{\kappa}\  - \ 1}\ \sum_{j = 1}^{n_{\kappa}}{(\lambda_{j}^{(\kappa)}\  - \ E_{\lambda}^{(\kappa)})^{2}}.\]

Each realization was analyzed independently. Across realizations, both
coordinates were summarized by their mean and sample standard deviation,
and each trajectory was represented by

\[({E\lambda}^{(\kappa)},\ {D\lambda}^{(\kappa)}),\ \ \ \ \ \kappa\  = \ 0,\ \ldots,\ 3,\]

in the mean--variance plane. For equal-length IID simulations,
between-realization SD represented finite-record dispersion of
individual coordinate estimates and was not divided by the square root
of the number of realizations. At each order, the 200 IID coordinates
were pooled across ten generators. The 95\% bivariate reference ellipse
was centered on the pooled mean and calculated from the sample
covariance matrix using the 95th percentile of the chi-square
distribution with two degrees of freedom; it represents finite-record
dispersion rather than uncertainty in the centroid.

\subsection*{1.2 Finite-record precision}

Finite-record precision was quantified from truncated gap sequences
extracted from continuous-IID inputs. At each order \ensuremath{\kappa} = 0--3,
\emph{E\textsubscript{\ensuremath{\lambda}}\textsuperscript{(\ensuremath{\kappa})}} and
\emph{D\textsubscript{\ensuremath{\lambda}}\textsuperscript{(\ensuremath{\kappa})}} were recalculated from
the first \emph{N\textsubscript{g}} = 250, 500, 1000, 2000, 5000 and
10000 gaps in each of 200 continuous-IID realizations spanning ten
generators. For each \emph{N\textsubscript{g}} and order, relative SD
was calculated across realizations as 100 \ensuremath{\times} SD / \textbar mean\textbar.
Values were averaged across hierarchy orders, with the order-wise
minimum and maximum retained to characterize between-order variation.
Unconstrained power-law fits were estimated by linear regression in
log--log coordinates. Cumulative estimates of the mean and unbiased
variance were also calculated from one representative IID realization
and compared with the corresponding complete-record values for
illustration only; these estimates were not used for record selection.
This calibration used prefixes of gap sequences extracted from long IID
records, rather than independent short input blocks. It therefore
quantified finite-gap-count precision without separately calibrating the
boundary effects of fragmented inputs.

\subsection*{1.3 Practical requirements and quality control}

Record length was planned from the deepest interpreted hierarchy order.
From the IID fits in Fig. 2, a target relative SD of \emph{r}\% requires
approximately \emph{N\textsubscript{g}} =
(36.5/\emph{r})\textsuperscript{2} gaps for \emph{E\textsubscript{\ensuremath{\lambda}}}
and \emph{N\textsubscript{g}} = (190/\emph{r})\textsuperscript{2} for
\emph{D\textsubscript{\ensuremath{\lambda}}}. Structured sequences may require more because
serial dependence alters effective sample size and extrema density. With
approximately one-third retention per extraction, 2,000 gaps at \ensuremath{\kappa} = 3
corresponds to about 1.5--1.6 \ensuremath{\times} 10\textsuperscript{5} initial
observations. This is a practical starting scale rather than a validity
threshold; shorter records may be used with quantified uncertainty, but
order-specific gap counts and the deepest supported order should be
reported.

\subsubsection*{Fragmented records: pooled precision and block-length adequacy}

Two distinct requirements apply when a record consists of separate
observation blocks. The pooled number of complete gaps controls sampling
precision, whereas the lengths of the individual blocks govern
finite-window effects. A gap is observed only when both bounding maxima
are retained within the same block; longer gaps are therefore
preferentially excluded by finite boundaries. Repeated extraction
shortens each block further, so an initially long block may become short
at deeper hierarchy orders. Pooling additional short blocks can narrow
the uncertainty around a displaced finite-block expectation without
removing that displacement.

Finite-length IID checks using the same asymmetric maxima rule indicated
that approximately 80 input values per independent block at the order
being analyzed kept the displacement of the mean and variance within
comparison tolerances of approximately 0.8\% and 3.8\%, respectively.
This is a provisional empirical guideline for the tested IID setting,
not a theoretical bound, a universal validity threshold or a guarantee
for structured inputs. These bias tolerances are distinct from the
across-realization relative SD fitted in Fig. 2. With roughly one-third
retention per extraction, 80 values entering order \(\kappa\) correspond
to approximately \(80\times3^{\kappa}\) original observations: about 80,
240, 720 and 2,160 for \(\kappa=0,1,2,3\), respectively. Actual retained
lengths should be assessed because extrema density depends on the
process. Neither a pooled count nor a mean block length establishes
adequacy of every contributing block.

For application-specific calibration, independent IID blocks should be
generated with the same input-length distribution as the observed
blocks. Maxima selection and re-indexing should be applied separately
within each block through the full hierarchy, and only completed
within-block gaps should be pooled at each order. Across repeated
ensembles, the pooled means and variances should be compared with the
corresponding long-series, order-specific IID reference. Varying the
number of blocks while holding their length distribution fixed separates
improved pooled precision from persistent finite-block displacement;
varying block length assesses the latter. Ensemble dispersion and
systematic displacement should be reported separately. This calibration
is additional to the gap-prefix precision analysis in Fig. 2, which does
not reproduce the repeated boundaries of many short input blocks.

For a temporal-order comparison, values should be shuffled independently
within each original block, preserving its length and empirical marginal
distribution, followed by identical block-wise hierarchy extraction.
The same outcome-independent inclusion rules should be used for data
and nulls. At each order, the numbers of contributing blocks and pooled
gaps, and the distribution of retained block lengths, should be reported.
Where block composition changes with order, a fixed-cohort sensitivity
analysis can distinguish hierarchy evolution from changing contributors.
Intervals, extrema and gaps should never be created across missing-data,
recording, subject or other independently defined block boundaries.
Matched-null comparisons remain informative when the long-series IID
reference is displaced by fragmentation, but matching does not make
short blocks asymptotically long or ensure adequate power at deep orders.

\subsubsection*{Other practical quality controls}

Slow trends and periodic components were treated as structure unless
independently identified as nuisance contamination; any removal should
be justified and checked against the unprocessed record. The present
controls did not validate linear detrending of temporally organized
signals. Continuously sampled data should resolve relevant fluctuations
without excessive acquisition smoothness dominating extrema selection;
stability under modest antialiased decimation provides a practical
check. Event timestamps should resolve the shortest relevant event.

Noise criteria were detector-specific. In binary-telegraph controls, raw
thresholding remained close to ground truth for SNR \textgreater{} 1; at
lower SNR, zero-phase fourth-order Butterworth filtering at 0.2 f\textsubscript{s} gave
the best tested compromise between false crossings and event mergers.
This cutoff is specific to the tested signal and detector. Filtering can
still merge short events and alter \ensuremath{\kappa}=0 statistics; higher-order
coordinates were more robust but were not assumed exact.

Finite amplitude resolution was tested in 20 independent U(0,1) IID
sequences of 10\textsuperscript{6} samples by quantizing values to fixed steps \ensuremath{\Delta}x = 10\textsuperscript{-6},
10\textsuperscript{-5}, 10\textsuperscript{-4}, 10\textsuperscript{-3}, 0.002, 0.003, 0.005, 0.007, 0.01, 0.1 and 0.2 before
applying the complete recursive analysis. At every order, the
adjacent-tie fraction was calculated as the number of adjacent
equal-valued pairs divided by the number of adjacent pairs, expressed as
a percentage. With the asymmetric plateau rule as described in Methods
§1.1, changes were negligible while the deepest-order tie fraction
remained about 3\% or less. At \ensuremath{\kappa} = 3, \ensuremath{\Delta}x = 0.002 produced 2.82\%
adjacent ties and shifted E\ensuremath{\lambda} and D\ensuremath{\lambda} by only 0.13\% and 0.64\%,
respectively, relative to unquantized controls. Systematic deviations
began around 4--5\% ties and became clear above 6--7\%, with D\ensuremath{\lambda} affected
more strongly than E\ensuremath{\lambda}. These values are empirical guides rather than
universal thresholds.

Structured-data sensitivity was tested using Mackey--Glass inter-peak
interval sequences at \ensuremath{\tau} = 18 and 30 (data generation is presented in
Methods, Sec. 1.5). For all 20 histories per delay, interval
values were canonicalized to the nominal 0.1-unit integration grid and
compared with the original sequences and independently shuffled matched
controls. Canonicalization changed organized-sequence coordinates by at
most 3.6 \ensuremath{\times} 10\textsuperscript{-5} in E\ensuremath{\lambda} and 1.1 \ensuremath{\times} 10\textsuperscript{-5} in D\ensuremath{\lambda}, while their order-dependent
displacement from shuffled controls was retained. The Mackey--Glass
trajectories were therefore not produced by finite-precision ties.

Quantized or discrete data therefore require an explicit tie rule and
order-specific reporting of adjacent-tie fractions. The continuous-IID
reference may be used when ties are negligible or have been shown
empirically not to affect the trajectory; otherwise, the reference
should preserve the marginal distribution, quantization and tie rule.
Missing blocks defined segment boundaries rather than being
concatenated. Data and nulls underwent identical sampling, filtering,
event detection, quantization and aggregation. For new applications,
record lengths should be matched where possible, uncertainty estimated
across independent realizations or segments, and the deepest interpreted
order specified in advance.

\subsection*{1.4 IID reference simulations and stochastic-model generators}

Unless stated otherwise, synthetic sequences contained
10\textsuperscript{6} samples and 20 independently initialized
realizations per condition. IID references comprised Gaussian N(0,1),
U(0,1), unit-scale exponential, lognormal (\ensuremath{\mu} = 0, \ensuremath{\sigma} = 1), gamma (shape
0.7, scale 1), and symmetric beta distributions (\ensuremath{\alpha} = \ensuremath{\beta} = 0.3, 0.5 or
0.8). Additional uniform and Gaussian controls used NumPy PCG64 and an
independent Box--Muller implementation. Permuted sinusoids, saw-tooth
waves and linear sequences preserved their empirical amplitude
distributions while removing serial order. Equal-length sources were
also added pointwise where indicated.

Gaussian autoregressive and moving-average controls were generated as

\[x_{t}\  = \ \varphi x_{t - 1}\  + \ \varepsilon_{t},\ \ \ \ \ \varepsilon_{t}\ \sim\ N(0,\ 1\  - \ \varphi^{2}).\]

\[x_{t}\  = \ \varepsilon_{t}\  + \ \sum_{j = 1}^{q}{\theta_{j}\varepsilon_{t - j}},\ \ \ \ \ \varepsilon_{t}\ \sim\ N(0,1).\]

respectively. AR(1) runs began at x\textsubscript{0} = 0, discarded 10\textsuperscript{4}
samples and used innovation scaling that maintained approximately unit
stationary variance; \emph{\ensuremath{\varphi}} and \emph{\ensuremath{\theta}} are reported with the
corresponding figures. A separate two-state Markov-modulated Gaussian
process was generated using symmetric transition probabilities
P\textsubscript{(A\ensuremath{\rightarrow}A)}=P\textsubscript{(B\ensuremath{\rightarrow}B)}=0.9 and
P\textsubscript{(A\ensuremath{\rightarrow}B)}=P\textsubscript{(B\ensuremath{\rightarrow}A)}=0.1. The hidden chain was
initialized in state A, and 10\textsuperscript{6} observations were drawn conditionally
from zero-mean Gaussian distributions with standard deviations
\ensuremath{\sigma}\textsubscript{A}=1.5 and \ensuremath{\sigma}\textsubscript{B}=0.3. The state was updated
after each observation.

In the systematic hidden-Markov Gaussian controls, variance switching
used \ensuremath{\mu}\textsubscript{A} = \ensuremath{\mu}\textsubscript{B} = 0, \ensuremath{\sigma}\textsubscript{A} =
0.5 and \ensuremath{\sigma}\textsubscript{B} = 1.5; mean switching used \ensuremath{\mu}\textsubscript{A}
= \ensuremath{-}1, \ensuremath{\mu}\textsubscript{B} = 1 and \ensuremath{\sigma}\textsubscript{A} = \ensuremath{\sigma}\textsubscript{B}
= 0.5. Symmetric state-persistence probabilities were 0.5, 0.6, 0.8 and
0.995. After a 10\textsuperscript{4}-sample burn-in, mean-switching
outputs were standardized.

Two nonlinear Gaussian controls were

\[x_{t}\  = \ \varepsilon_{t}\varepsilon_{t - 1},\]

\[x_{t}\  = \ \varepsilon_{t}\  + \ 0.5(\varepsilon_{t - 1}^{2}\  - \ 1),\]

The first was variance-normalized and the second standardized to zero
mean and unit variance.

\subsection*{1.5 Deterministic and chaotic data generators}

Twenty trajectories were generated for each discrete map after 1,000
transient iterations. The logistic map used \emph{r} = 4 and
x\textsubscript{0} = 0.172, 0.173, \ldots, 0.191:

\[x_{n+1}\  = \ r\ x_{n}(1\  - \ x_{n}),\]

The tent map was x\textsubscript{n+1} = ax\textsubscript{n} for
x\textsubscript{n} \textless{} 0.5 and \emph{a}(1 \ensuremath{-} x\textsubscript{n})
otherwise, with \emph{a} = 1.999 and the same initial values as the
logistic map. The Hénon map used \emph{a} = 1.4, \emph{b} = 0.3,
y\textsubscript{0} = 0 and x\textsubscript{0} = 0.051, 0.052, \ldots,
0.070:

\[x_{n+1}\  = \ 1\  - \ ax_{n}^{2}\  + \ y_{n},\ \ \ \ \ \ y_{n+1}\  = \ bx_{n},\]

Only its \emph{x}-coordinate was analyzed.

The chaotic Chua circuit was simulated as

\[\dot{x}{}\  = \ \alpha\lbrack y\  - \ x\  - \ h(x)\rbrack,\ \ \ \ \ \ \dot{y}{}\  = \ x\  - \ y\  + \ z,\ \ \ \ \ \ \dot{z}{}\  = \  - \beta y,\]

with

\[h(x)\  = \ m_{1}x\  + \ \tfrac{1}{2}{}(m_{0}\  - \ m_{1})(|x\  + \ 1|\  - \ |x\  - \ 1|).\]

where \ensuremath{\alpha} = 15.6, \ensuremath{\beta} = 28.0, m\textsubscript{0} = \ensuremath{-}1.143 and
m\textsubscript{1} = \ensuremath{-}0.714. Fourth-order Runge--Kutta integration used
\emph{dt} = 0.01 for 2 \ensuremath{\times} 10\textsuperscript{8} steps, discarding
10\textsuperscript{5}. Twenty trajectories began at
(\emph{x\textsubscript{0},0,0}), with \emph{x\textsubscript{0}} = 0.075,
0.077, \ldots, 0.113. Positive and negative \emph{x} dwells were
delimited by hysteretic thresholds \ensuremath{-}0.018 and +0.018; positive, negative
and alternating dwell sequences were analyzed separately, with durations
expressed as samples \ensuremath{\times} \emph{dt}. The final complete dwell was included.

The Mackey--Glass equation was

\[dx/dt\  = \ \beta x(t\  - \ \tau)/\lbrack 1\  + \ x(t\  - \ \tau)^{n}\rbrack\  - \ \gamma x(t),\]

with \emph{\ensuremath{\beta}} = 0.2, \emph{\ensuremath{\gamma}} = 0.1, n = 10, \emph{dt} = 0.1 and \ensuremath{\tau} = 18,
30 or 60. Fourth-order Runge--Kutta integration used linear
interpolation of delayed intermediate-stage values. Runs contained
10\textsuperscript{8} steps after 50,000 transients; 20 constant
histories used \emph{x\textsubscript{0}} = 1.20, 1.21, \ldots, 1.39.
Peaks followed the plateau convention above, and inter-peak intervals in
physical time were analyzed.

The nonlinear four-dimensional electronic spiking neuron was simulated
in the TINA\texttrademark{} SPICE environment. The schematic is available in our
previous publication\textsuperscript{18}. For each simulation, 100,000 s
of activity were generated at 100 Hz and the first 60 s were removed
before spike extraction. Spike times were defined by the first upward
crossing of the voltage threshold V\textsubscript{threshold}=1.0 V. For
each crossing, the threshold-crossing time was estimated by linear
interpolation between the two consecutive samples bracketing the
threshold, using their recorded time and voltage coordinates.
Inter-spike intervals (ISIs) were calculated as differences between
successive interpolated crossing times; therefore, ISI values were not
restricted to integer multiples of the sampling interval \emph{dt}=0.01
s.

\subsection*{1.6 Signal transformations and nonstationary controls}

Gaussian IID sequences were smoothed with a normalized rectangular
moving average of length \emph{L},

\[y_{i}\  = \ (1/L)\sum_{_{j=0}}^{^{L-1}}\ \ x_{i+j}.\]

using complete windows displaced by one sample. Separately, Keras-style
windows of length \emph{L}, sampling rate 1 and stride \emph{S} were
reduced to their mean, maximum or last sample; outputs were standardized
before analysis. Stride was varied to control window overlap.

Periodic contamination was generated as

\[y_{i}\  = \ x_{i}\  + \ A\ sin(2\pi i/P\  + \ \varphi),\]

where \emph{A} set the sinusoid-to-source standard-deviation ratio,
\emph{P} its period and \emph{\ensuremath{\varphi}} a phase fixed within each simulation.
For the tested decontamination, the sequence mean was removed, the
dominant FFT frequency was sought over periods of 20--20,000 samples,
and the estimate was refined over 400 frequencies spanning \ensuremath{\pm}2 FFT bins.
At each frequency, \emph{c + a cos(2\ensuremath{\pi}fi) + b sin(2\ensuremath{\pi}fi)} was fitted by
least squares; the minimum-residual fit was selected, and only its
oscillatory component was subtracted. The cleaned and contaminated
sequences were analyzed identically.

Linear non-stationarity was introduced as

\[x_{i}(trend)\  = \ x_{i}\  + \ c\ \sigma _{x}\ i,\]

where \emph{c} set signed trend strength relative to source standard
deviation \ensuremath{\sigma}\textsubscript{x}. Linear detrending was not tested.
Independent exponential samples with a deterministic scale transition
used

\[\tau(i)\  = \ \tau _{1}\  + \ (\tau _{2}\  - \ \tau _{1})/\{ 1\  + \ exp\lbrack - (i\  - \ i_{0})/W\rbrack\},\]

and

\[x_{i}\ \sim\ Exp\lbrack\tau(i)\rbrack.\]

where \emph{\ensuremath{\tau}\textsubscript{1}} and \emph{\ensuremath{\tau}\textsubscript{2}} are pre- and post-transition scales,
\emph{i\textsubscript{0}} is the transition location and \emph{W} its width.
Condition-specific \emph{L, S, A, P, c, \ensuremath{\tau}} and \emph{W} values are
reported with the corresponding results.

\subsection*{1.7 Event-derived dwell sequences and membrane-memory models}

For event-derived data, the analyzed sequence comprised dwell durations, inter-peak intervals or inter-spike intervals.

Binary trajectories alternated between closed (0) and open (1) states.
Open dwells, closed dwells and the complete alternating dwell sequence
were analyzed separately. IID dwell durations were geometric with

\[p\  = \ 1/\tau,\]

and draws shorter than two samples were rejected. Equal or unequal
state-specific scales generated symmetric or asymmetric controls without
dependence within either state.

Memory was introduced through a hidden environment over successive
dwells,

\[m_{i} = \rho_{env}m_{i - 1}\  + \sigma_{env}\eta_{i},\]

which was standardized; unless stated otherwise, \ensuremath{\rho}\textsubscript{env} =
0.995 and \ensuremath{\sigma}\textsubscript{env} = 0.08. Open and closed scales were
oppositely coupled to it:

\[\tau_{i}^{open}\  = \ \tau_{open,0}\ exp(\alpha m_{i}\  - \ \alpha ^{2}/2),\]

\[\tau_{i}^{closed}\  = \ \tau_{closed,0}\ exp( - \alpha m_{i}\  - \ \alpha ^{2}/2),\]

The \ensuremath{-}\ensuremath{\alpha}\textsuperscript{2}/2 term preserved the nominal mean scale for a standardized
Gaussian environment. Scales were restricted to 1.000001--10\textsuperscript{6}; \ensuremath{\alpha}
controlled memory strength independently of the baseline asymmetry
\ensuremath{\tau}\textsubscript{open,0}/\ensuremath{\tau}\textsubscript{closed,0}.

Generated dwells were also concatenated into binary traces of up to
10\textsuperscript{8} samples; only complete events were retained, with
the matched ground-truth subset stored. Independent Gaussian measurement
noise of standard deviation \ensuremath{\sigma}\textsubscript{noise} was added, with SNR =
1/(6.6\ensuremath{\sigma}\textsubscript{noise}) for the unit-amplitude signal. At
\emph{f\textsubscript{s}} = 100 Hz, filtering used a fourth-order
Butterworth low-pass filter applied forward and backward; cutoffs were
expressed relative to \emph{f\textsubscript{s}}.

Traces were thresholded at 0.5 and run-length encoded. Runs shorter than
two samples were treated as crossing artifacts: internal runs bounded by
the same state were merged with both neighbors, and boundary runs with
their sole neighbor. An independent IID telegraph implementation
underwent the same trace generation, noise, filtering and extraction
pipeline as a matched processing control.

\subsection*{1.8 Synthetic neuronal population generators}

Five independent synthetic populations were generated per model
condition. Each contained 50 neurons, grouped into ten non-overlapping
five-neuron clusters and one shank, with 150,000 inter-spike intervals
(ISIs) per neuron. Spike times used a 20-kHz integer grid (0.05 ms) and
a 5-ms absolute refractory period. Baseline rate was 1.33 Hz;
neuron-specific rates were lognormally distributed with coefficient of
variation 0.10, and initial offsets were drawn up to 25 ms.

For AR-modulated populations, 50-ms-grid rate fields were

\[z_{i}(t)\  = \ \sqrt{c}\ G(t)\  + \ \sqrt{(1 - c)}\ E_{i}(t),\]

where \emph{G} and \emph{E\textsubscript{i}} were population-wide and
neuron-specific AR(1) components with coefficients 0.995 and 0.990.
Rates were

\[r_{i}(t)\  = \ r_{0,i}\ exp\lbrack\beta z_{i}(t)\  - \ \beta ^{2}/2\rbrack,\]

with \emph{\ensuremath{\beta}} = 1.60 and \emph{c} = 0 or 0.75; \emph{c} = 0.75
represented the sleep-like condition with predominantly shared
population modulation. Spikes were drawn as refractory point processes.
A burst condition used the same baseline trains: eligible spikes
initiated a proposed 10-spike burst with probability 0.097, with nine
intervals sampled uniformly from 10--20 ms. Bursts were accepted only if
they ended before the next baseline spike and respected the refractory
period, yielding approximately 5\% realized enrichment.

Two shared hidden-state models used symmetric 50-ms-bin Markov states
with P\textsubscript{stay} = 0.9996666667 (mean duration \ensuremath{\approx}150 s). The
mean-switching model selected rates 0.67 or 2.00 Hz with
shifted-exponential ISIs; the variance-switching model held expected ISI
constant while changing shifted-Gamma shape from 5.0 to 0.5. Neurons
were conditionally independent given the state.

Multiscale modulation combined seven stationary AR(1) components with
timescales 0.1, 0.4, 1.6, 6.4, 25.6, 102.4 and 409.6 s. Components were
equally weighted additively or arranged in a nested multiplicative
cascade. The shared standardized field modulated neuron-specific
baselines exponentially, with \emph{\ensuremath{\beta}} = 1.20 and cascade gating
strength 0.55.

Five spike trains were truncated to common duration, concatenated and
sorted to form each cluster; ten clusters were superposed similarly into
a shank. Events sharing a grid timestamp were represented once and
mergers counted. A paired shank-level resolution control superposed the
same floating-point source times continuously or after 20-kHz
quantization and identical-timestamp collapse. For multiscale shanks,
maximum absolute paired differences were 0.0098 in \emph{E\ensuremath{\lambda}} and 0.0242
in \emph{D\ensuremath{\lambda}}, without changing hierarchy classification.

For each organized population, a source-shuffled null independently
permuted each neuron's ISIs while preserving its multiset, count and
initial timestamp, followed by the same cluster and shank superposition.
A second shank null permuted ISIs within each already-organized cluster,
preserving its ISI multiset, count, endpoints and duration before shank
superposition. This tested dependence on the level at which serial order
was removed.

\subsection*{1.9 Experimental neuronal population recordings}

Extracellular recordings from freely moving mice were obtained from the
public Buzsáki Laboratory th-1 dataset in the NYU Langone Health
DTN1--Buzsakilab collection (Globus). Spike timestamps and assignments
came from Klusters/Neuroscope .res.N and .clu.N files; XML metadata
specified 20-kHz acquisition. Cluster IDs 0 and 1 were excluded, other
curated clusters retained, and shanks with fewer than two accepted
clusters omitted. Day numbers denote chronologically ordered sessions.

Analysis was restricted to slow-wave-sleep (SWS) intervals. Within each
interval, accepted shank-cluster timestamps were pooled and sorted; no
coincident retained timestamps occurred. Coincidences removed during
original sorting or curation cannot be recovered from the public files.
ISIs were formed within individual SWS bouts. For the normalized
separation analysis (Fig. \hyperref[fig:6]{6}C), maxima were extracted
and re-indexed independently within each bout at every order.
Successive-maxima gaps were calculated within bouts and then pooled by
day and shank to estimate the mean and sample variance at each order.

A strict-local null was generated for each day \ensuremath{\times} shank by permuting ISIs
within each cluster and SWS interval while fixing the first spike, then
reconstructing the train by cumulative summation. This preserved the
interval-specific ISI multiset, spike count, endpoints, occupancy, local
rate and SWS boundaries while removing serial order. Reconstructed
clusters were superposed as for recorded data.

Five deterministic replicate-specific shuffles were generated per
realization and processed identically to recorded sequences.
Order-specific analyses required finite \emph{E\ensuremath{\lambda}} and \emph{D\ensuremath{\lambda}} for the
recorded sequence and all five nulls; complete-trajectory analyses
required paired coordinates through \ensuremath{\kappa}=3. Pairing retained mouse, day,
shank, constituent clusters and SWS intervals.

\subsection*{1.10 Paired residuals, normalized separation and regime comparisons}

Organized-minus-null residuals were calculated in the mean--variance
plane at each hierarchy order:

\(\Delta E\lambda^{(\kappa)} = E\lambda_{org}^{(\kappa)} - E\lambda_{null}^{(\kappa)}\)

and

\[\Delta D\lambda^{(\kappa)} = D\lambda_{org}^{(\kappa)} - D\lambda_{null}^{(\kappa)}\]

Synthetic observations were paired within model, population realization
and neuron, cluster or shank identity before subtraction. Each model and
order yielded 250 neuron, 50 cluster and 5 shank pairs, summarized by
mean and sample SD. The source-shuffled lineage was the primary
reference; the cluster-shuffled shank control was summarized separately.

For experimental recordings, the null coordinate was the centroid of the
five strict-local nulls for the same day \ensuremath{\times} shank. Because these
coordinates formed correlated, approximately elliptical clouds in the
mean--variance plane, separation used an order-specific Mahalanobis
metric.

For each day \ensuremath{\times} shank record and hierarchy order \(\kappa\), let the
recorded coordinate be

\[x_{real}^{(\kappa)} = \begin{pmatrix}
E\lambda_{real}^{(\kappa)} \\
D\lambda_{real}^{(\kappa)}
\end{pmatrix},
\]and let the five matched shuffled-source null coordinates be
\(x_{r}^{(\kappa)}\), \(r = 1,\ldots,5\). Their centroid is

\[{\bar{x}}_{null}^{(\kappa)} = \frac{1}{5}\sum_{r = 1}^{5}x_{r}^{(\kappa)}.
\]The real-minus-null residual vector is

\[\Delta^{(\kappa)} = x_{real}^{(\kappa)} - {\bar{x}}_{null}^{(\kappa)}.
\]For the Mahalanobis metric, the order-specific covariance matrix was
estimated by pooling within-shank deviations of the five null
realizations from their shank-specific centroids across the 38 eligible
day \ensuremath{\times} shank records. The real-to-null Mahalanobis distance is

\[R_{real}^{(\kappa)} = \sqrt{\left( \Delta^{(\kappa)} \right)^{T}\left( \Sigma^{(\kappa)} \right)^{- 1}\Delta^{(\kappa)}}.
\]For each null realization \(r\), define the leave-one-out centroid of
the other four nulls as

\[{\bar{x}}_{- r}^{(\kappa)} = \frac{1}{4}\sum_{\substack{s = 1 \\ s \neq r}}^{5}x_{s}^{(\kappa)}.
\]Its Mahalanobis distance from that leave-one-out centroid is

\[d_{r}^{(\kappa)} = \sqrt{\left( x_{r}^{(\kappa)} - {\bar{x}}_{- r}^{(\kappa)} \right)^{T}\left( \Sigma^{(\kappa)} \right)^{- 1}\left( x_{r}^{(\kappa)} - {\bar{x}}_{- r}^{(\kappa)} \right)}.
\]Finally,

\[Q^{(\kappa)} = \frac{R_{real}^{(\kappa)}}{\operatorname*{median}_{r = 1,\ldots,5}{(d_{r}^{(\kappa)})}}\]

Thus Q = 1 denotes real--null separation equal to the typical
null-to-null scale in the covariance-normalized plane. Q was used as a
descriptive separation ratio, not as a record-level probability or
significance test. The same 38 records were used at all four orders. Eligibility for
Fig. \hyperref[fig:6]{6}C required at least 2,000 pooled within-bout gaps
at \ensuremath{\kappa}=3 separately for the recording and each of its five nulls.
Distributions across records were summarized by medians and interquartile ranges.

Two dynamic regimes can be compared directly or relative to their own
nulls. Direct coordinate subtraction measures total displacement at each
order. With matched nulls, subtracting each null coordinate before
comparing residual trajectories isolates changes in temporal ordering
after regime-specific marginals and processing are accounted for.
Pairing should be retained across subjects, units or realizations, with
uncertainty estimated from paired differences.

\section*{Competing interests}
No competing interests are disclosed.

\section*{Acknowledgments}

The author gratefully acknowledges the late Gleb Evlampievich Glazirin
(1939--2021), whose idea of recursive extrema-gap statistics provided
the starting point for this way of looking at real-world data. The
author thanks Giorgio Sonnino and Pasquale Nardone for a careful and
critical reading of the manuscript.

\onecolumn
\section*{Appendix. Supplementary figures}
\setcounter{figure}{0}
\renewcommand{\theHfigure}{appendix.\arabic{figure}}
\renewcommand{\thefigure}{S\arabic{figure}}

\begin{figure}[!htp]
\centering
\includegraphics[width=0.92\textwidth,height=7.0in,keepaspectratio]{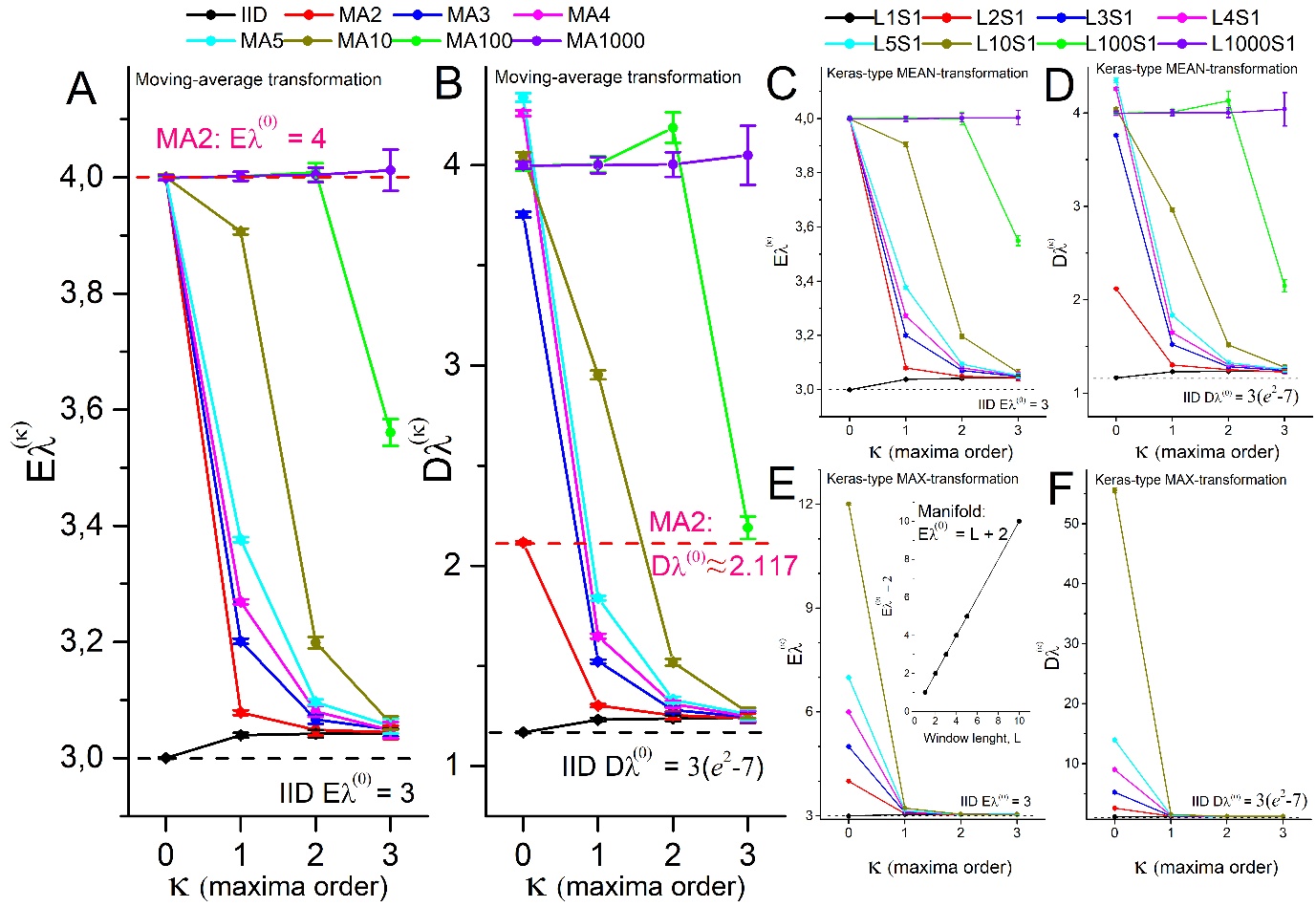}
\caption{\textbf{Effects of overlapping smoothing and window-based
transformations on hierarchy statistics.}\\
\textbf{A,B,} Mean gap length E\ensuremath{\lambda}\textsuperscript{(\ensuremath{\kappa})} and gap variance
D\ensuremath{\lambda}\textsuperscript{(\ensuremath{\kappa})} for Gaussian IID sequences subjected to
rectangular moving averages of the indicated lengths. The unsmoothed IID
control is shown in black. Dashed lines mark the exact raw IID values
(E\ensuremath{\lambda}\textsuperscript{(0)}=3) and
(D\ensuremath{\lambda}\textsuperscript{(0)}=3(e\textsuperscript{2}-7)); the analytical
two-point moving-average results (E\ensuremath{\lambda}\textsuperscript{(0)}=4 and
D\ensuremath{\lambda}\textsuperscript{(0)}\ensuremath{\approx}2.117) are also indicated. \textbf{C,D,}
Corresponding results after Keras-type mean reduction of overlapping
windows of length (L) and stride (S=1). \textbf{E,F,} Results after
maximum reduction of the same windows. The inset in E shows the
raw-order relation E\ensuremath{\lambda}\textsuperscript{(0)}=L+2. In all transformations,
the largest displacement occurs at low hierarchy orders and decreases
under recursive extraction. Points and error bars are mean \ensuremath{\pm} SD across
20 independent realizations per condition; where not visible, error bars
are smaller than the symbols.}
\label{fig:S1}
\end{figure}

\clearpage

\begin{figure}[!htp]
\centering
\includegraphics[width=0.92\textwidth,height=6.7in,keepaspectratio]{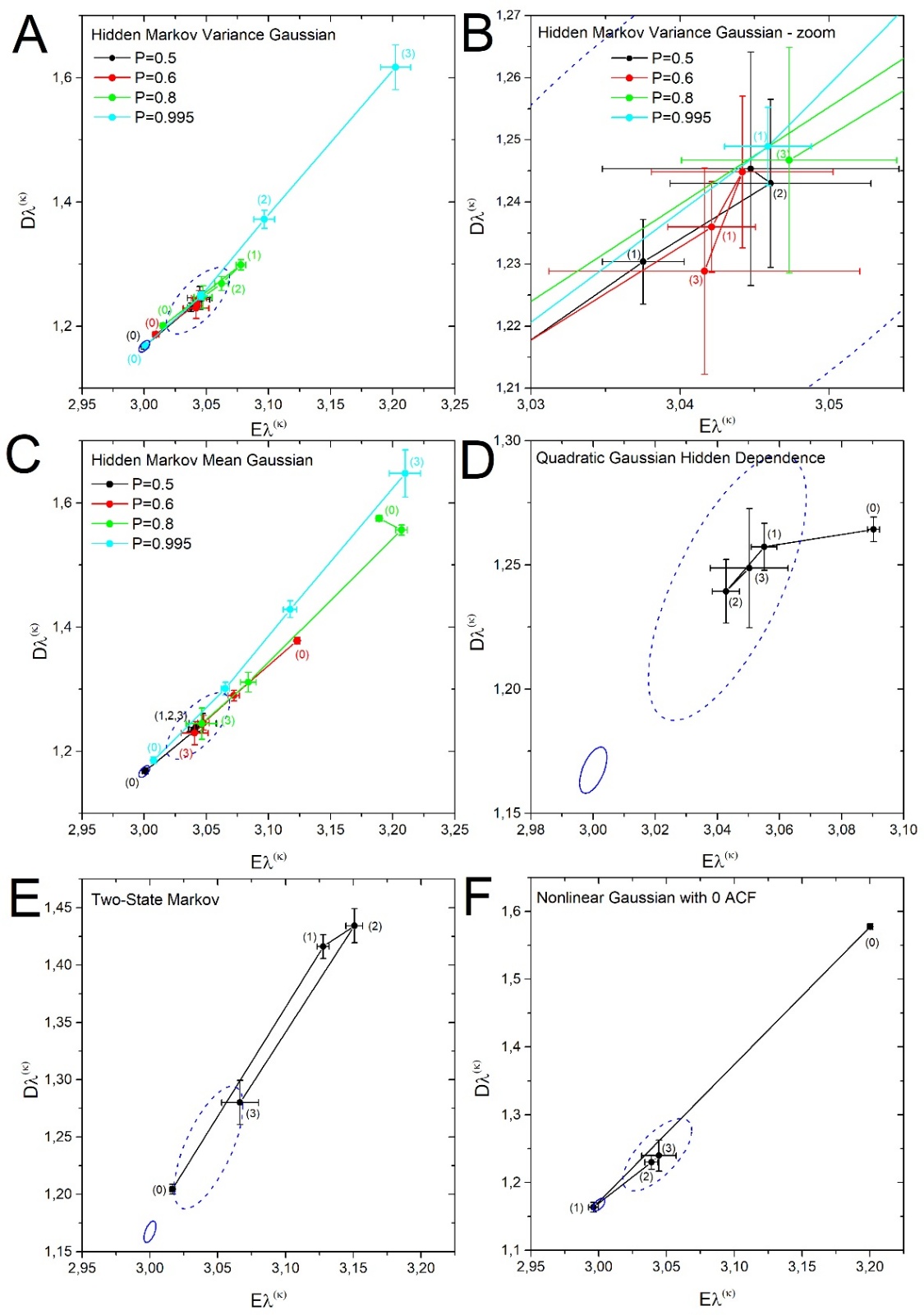}
\caption{\textbf{Hierarchy trajectories generated by hidden-state and
nonlinear stochastic dependence.}\\
\textbf{A,} Variance-switching Gaussian hidden-Markov processes with
symmetric state-persistence probabilities P\textsubscript{stay}=0.5,
0.6, 0.8 and 0.995. \textbf{B,} Enlargement of the region containing the
lower-persistence trajectories in A. \textbf{C,} Corresponding
mean-switching Gaussian hidden-Markov processes. \textbf{D,} Gaussian
process with hidden quadratic lag dependence,
\(x_{t}\  = \ \varepsilon_{t}\  + \ 0.5\left( \varepsilon_{t - 1}^{2}\  - \ 1 \right).\ \)\textbf{E,}
Two-state Markov-modulated Gaussian process with symmetric
state-persistence probability 0.9 and state-dependent standard
deviations 1.5 and 0.3. \textbf{F,} Variance-normalized nonlinear
Gaussian process \(x_{t}\  = \ \varepsilon_{t}\varepsilon_{t - 1},\)
which has zero linear autocorrelation at non-zero lags. Numbers indicate
hierarchy orders. Points and error bars are mean \ensuremath{\pm} SD across 20
independent realizations. Solid and dashed blue ellipses denote the 95\%
IID reference regions at \ensuremath{\kappa}=0 and \ensuremath{\kappa}=3, respectively.}
\label{fig:S2}
\end{figure}

\clearpage

\begin{figure}[!htp]
\centering
\includegraphics[width=0.92\textwidth,height=7.0in,keepaspectratio]{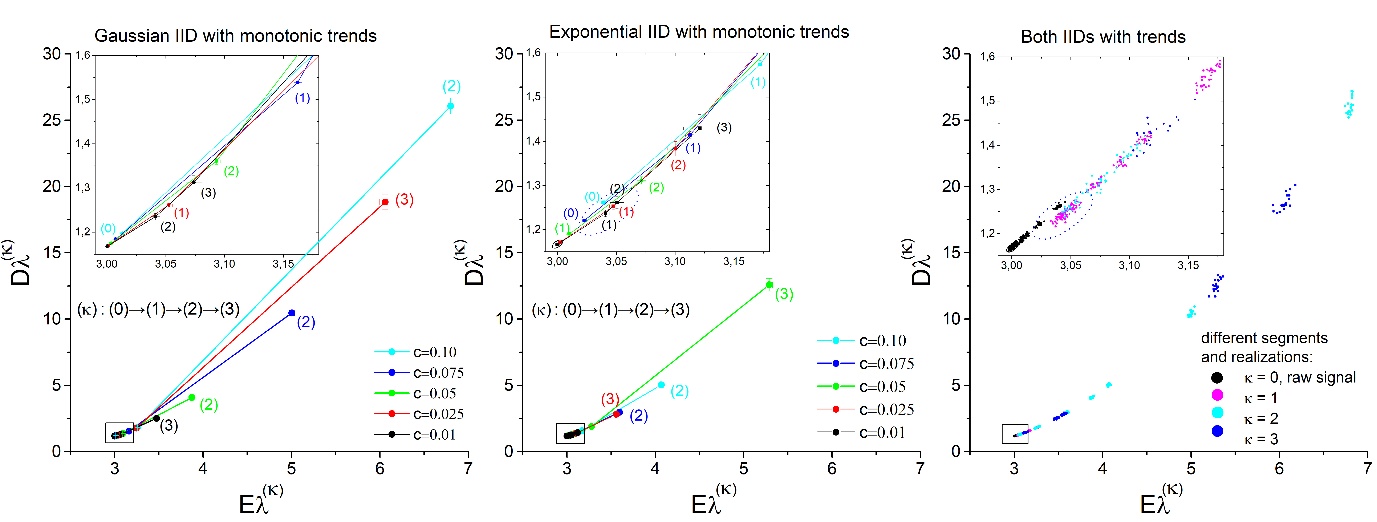}
\caption{\textbf{Hierarchy amplification induced by monotonic
trends.}\\
\textbf{A,} Hierarchy trajectories of Gaussian IID sequences after
addition of positive monotonic trends with normalized slopes
(c=0.01--0.10). \textbf{B,} Corresponding trajectories for exponential
IID sequences. Numbers identify hierarchy orders and lines connect k~:
(0) \ensuremath{\rightarrow} (1) \ensuremath{\rightarrow} (2) \ensuremath{\rightarrow} (3); insets enlarge the weak-trend region. \textbf{C,}
Combined results from different segments and realizations of the
Gaussian and exponential data, coloured by hierarchy order. Increasing
trend strength displaces the trajectories progressively from the IID
region, with broadly similar geometry for the two amplitude
distributions. Points in A and B are means across 20 independent
realizations per condition; error bars indicate SD and are generally
smaller than the symbols.}
\label{fig:S3}
\end{figure}

\clearpage

\begin{figure}[!htp]
\centering
\includegraphics[width=0.49\textwidth,height=7.0in,keepaspectratio]{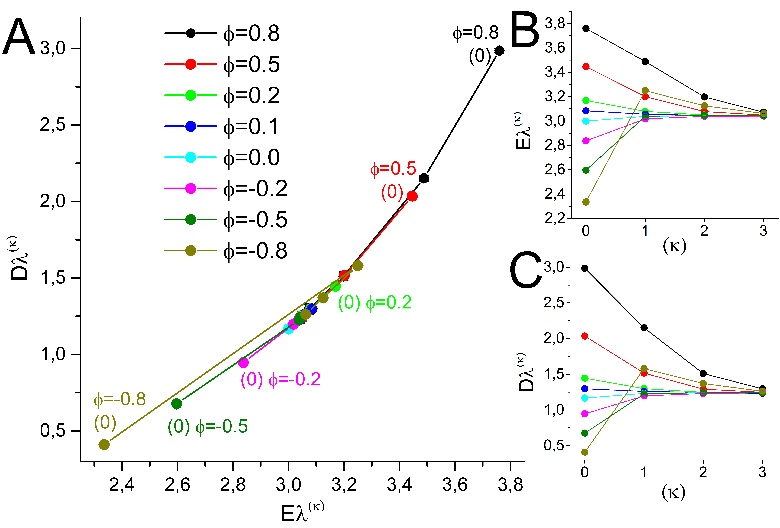}
\includegraphics[width=0.49\textwidth,height=7.0in,keepaspectratio]{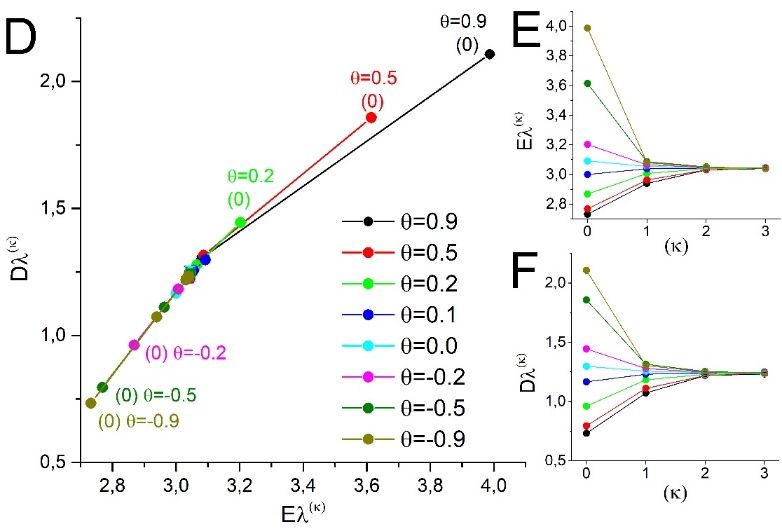}
\caption{\textbf{Distinct hierarchy manifolds generated by
autoregressive and moving-average dependence.}\\
\textbf{A,} Mean--variance trajectories of Gaussian AR(1) processes with
autoregressive coefficients \ensuremath{\phi}=-0.8 to 0.8. \textbf{B,C,} Corresponding
evolution of E\ensuremath{\lambda}\textsuperscript{(\ensuremath{\kappa})} and D\ensuremath{\lambda}\textsuperscript{(\ensuremath{\kappa})} with
hierarchy order. \textbf{D,} Mean--variance trajectories of one-lag
Gaussian moving-average processes with coefficients (\(\theta\) =-0.9 to
0.9). \textbf{E,F,} Corresponding evolution of the two gap statistics.
The \(\varphi\)=0 and \(\theta\)=0 conditions are IID controls. AR and
moving-average dependence generate oppositely curved manifolds at low
hierarchy orders, whereas their trajectories approach a common
higher-order region. Points and error bars are mean \ensuremath{\pm} SD across 20
independent realizations per coefficient; where not visible, error bars
are smaller than the symbols.}
\label{fig:S4}
\end{figure}

\clearpage

\begin{figure}[!htp]
\centering
\includegraphics[width=0.92\textwidth,height=7.0in,keepaspectratio]{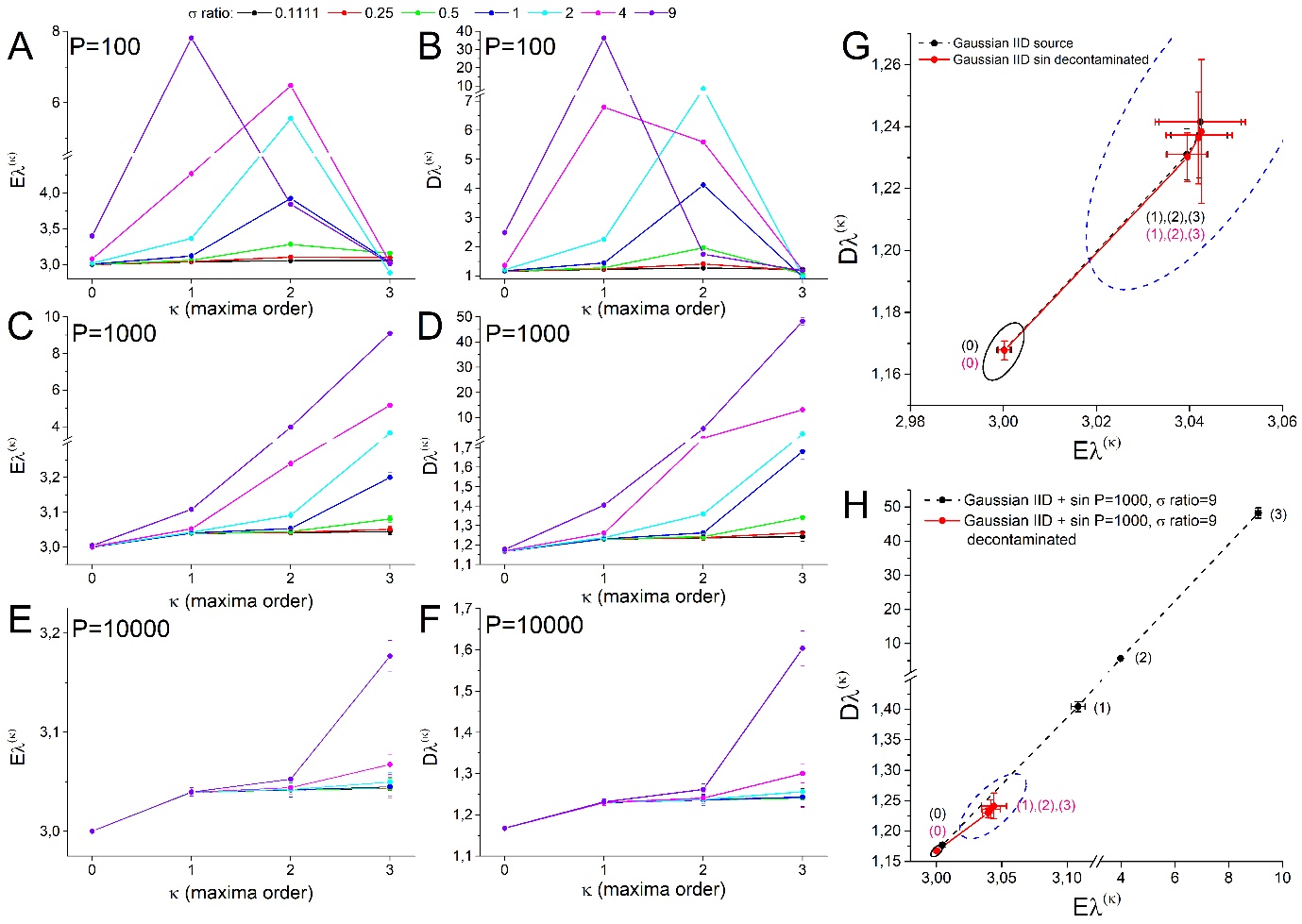}
\caption{\textbf{Dependence of periodic-contamination effects on
temporal scale and amplitude.}\\
\textbf{A,B,} E\ensuremath{\lambda}\textsuperscript{(\ensuremath{\kappa})} and D\ensuremath{\lambda}\textsuperscript{(\ensuremath{\kappa})} for
Gaussian IID sequences contaminated by sinusoids of period (P=100)
samples. Colours denote sinusoid-to-source standard-deviation ratios
from 0.1111 to 9. \textbf{C,D,} Corresponding results for (P=1000).
\textbf{E,F,} Corresponding results for (P=10000). The hierarchy
excursion depends jointly on contaminant period and relative amplitude.
\textbf{G,} Gaussian IID source trajectories before and after
application of the sinusoid-fitting and subtraction procedure in the
absence of an added sinusoid. \textbf{H,} Trajectories before and after
fitted-sinusoid removal for the strongly contaminated condition
(P=1000), standard-deviation ratio 9. The fitted removal restores the
trajectory to the empirical IID region. This decontamination test was
restricted to sinusoid-contaminated Gaussian IID inputs. Points and
error bars are mean \ensuremath{\pm} SD across 20 independent realizations. Solid and
dashed ellipses denote the 95\% IID reference regions at \ensuremath{\kappa}=0 and \ensuremath{\kappa}=3,
respectively.}
\label{fig:S5}
\end{figure}

\clearpage

\begin{figure}[!htp]
\centering
\includegraphics[width=0.92\textwidth,height=7.0in,keepaspectratio]{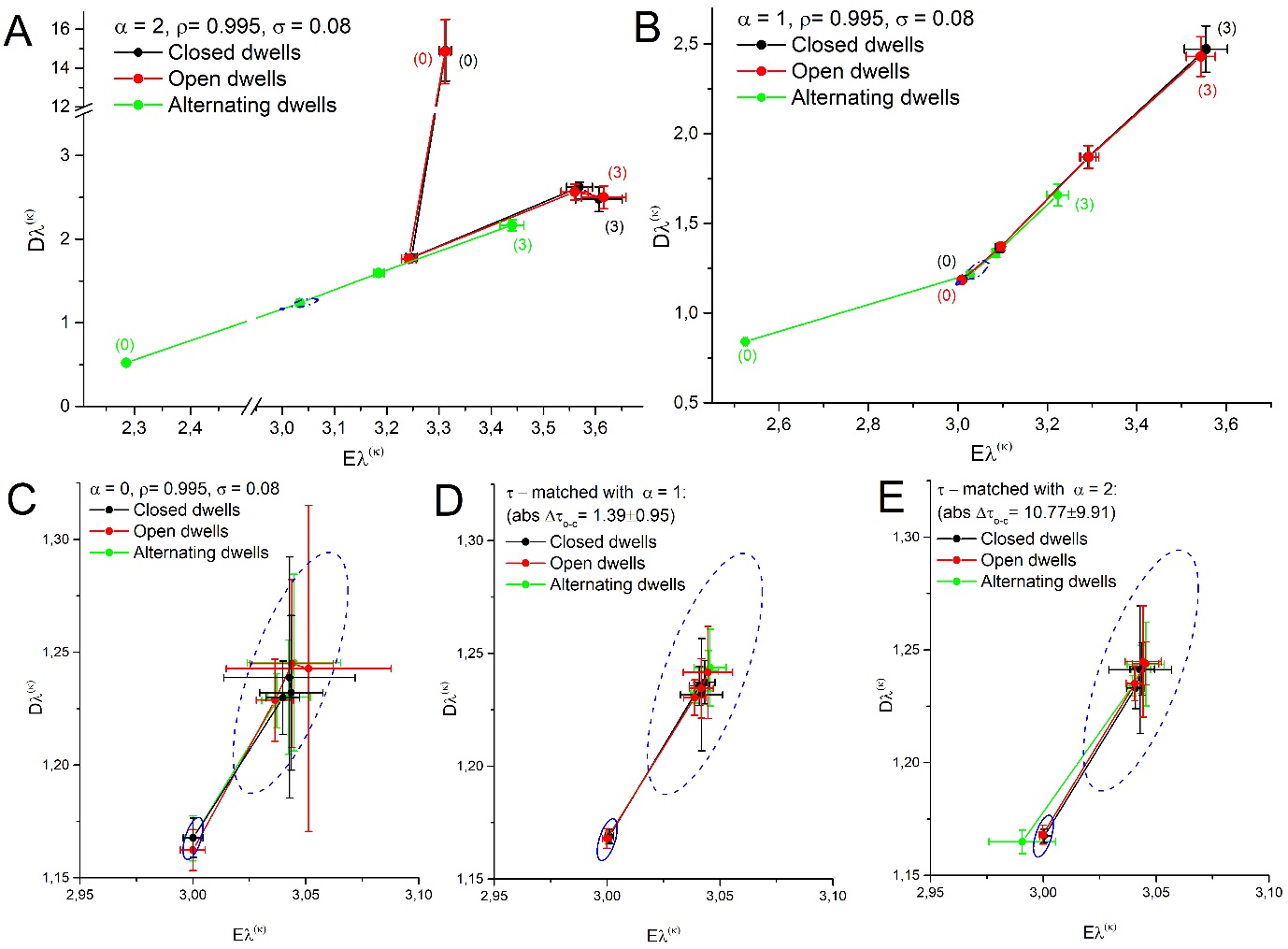}
\caption{\textbf{Environmental memory produces persistent dwell-time
hierarchy displacement.}\\
\textbf{A,} Closed, open and alternating dwell trajectories for strong
environmental coupling \ensuremath{\alpha}=2, \ensuremath{\rho}=0.995, \ensuremath{\sigma}=0.08. \textbf{B,} Corresponding
trajectories for weaker coupling \ensuremath{\alpha}=1. \textbf{C,} Memory-free condition
\ensuremath{\alpha}=0 with the same environmental-process parameters. \textbf{D,E,}
Independent dwell controls whose open and closed mean durations were
matched to those generated with \ensuremath{\alpha}=1 and \ensuremath{\alpha}=2, respectively. The mean
absolute open--closed duration differences were 1.39\ensuremath{\pm}0.95 and 10.77\ensuremath{\pm}9.91
sampling intervals. These matched controls separate temporal memory from
differences in marginal dwell duration. Black, red and green denote
closed, open and alternating dwells, respectively; numbers identify
hierarchy orders. Points and error bars are mean \ensuremath{\pm} SD across 20
independent realizations. Solid and dashed blue ellipses denote the 95\%
IID reference regions at \ensuremath{\kappa}=0 and \ensuremath{\kappa}=3.}
\label{fig:S6}
\end{figure}

\clearpage

\begin{figure}[!htp]
\centering
\includegraphics[width=0.92\textwidth,height=7.0in,keepaspectratio]{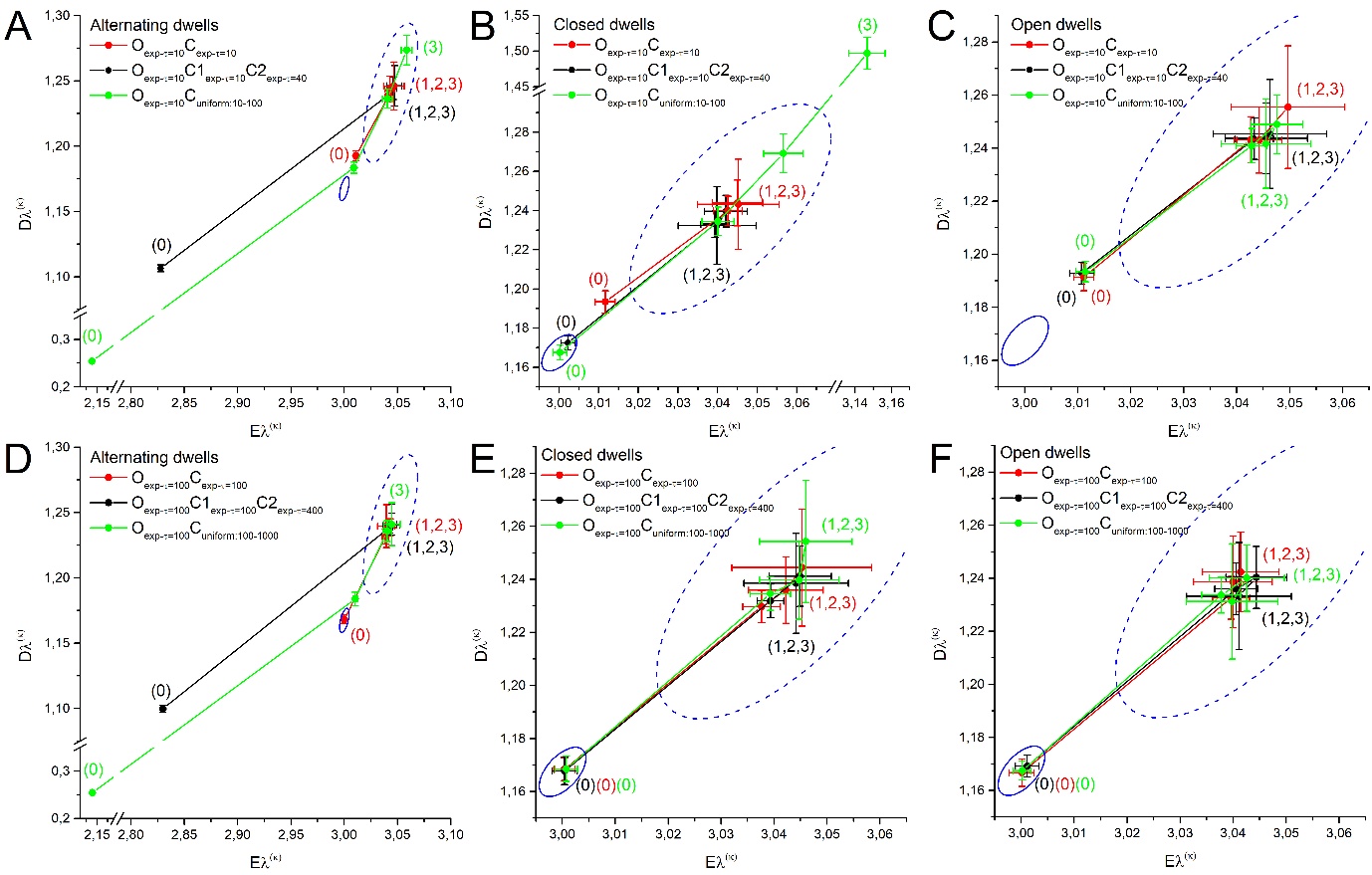}
\caption{\textbf{Effects of event alternation, temporal resolution and
dwell-distribution heterogeneity in memory-free sequences. A--C,}
Hierarchy trajectories of alternating, closed and open dwells when the
characteristic exponential dwell scale is approximately ten sampling
intervals. Red denotes equal-scale open and closed dwells, black denotes
an exponential open state combined with two exponential closed-state
components, and green denotes an exponential open state combined with a
broad uniform closed-dwell distribution; exact parameters are given in
the panel legends. \textbf{D--F,} Corresponding analyses after
increasing the characteristic scales approximately tenfold. Raw
alternating-dwell statistics can be strongly displaced when events
approach the sampling scale or when the two states have different
distributions. Higher hierarchy orders nevertheless move toward the
corresponding IID regions in these memory-free controls. Points and
error bars are mean \ensuremath{\pm} SD across 20 independent realizations. Solid and
dashed blue ellipses denote the 95\% IID reference regions at \ensuremath{\kappa}=0 and
\ensuremath{\kappa}=3, respectively.}
\label{fig:S7}
\end{figure}

\clearpage

\begin{figure}[!htp]
\centering
\includegraphics[width=0.92\textwidth,height=7.0in,keepaspectratio]{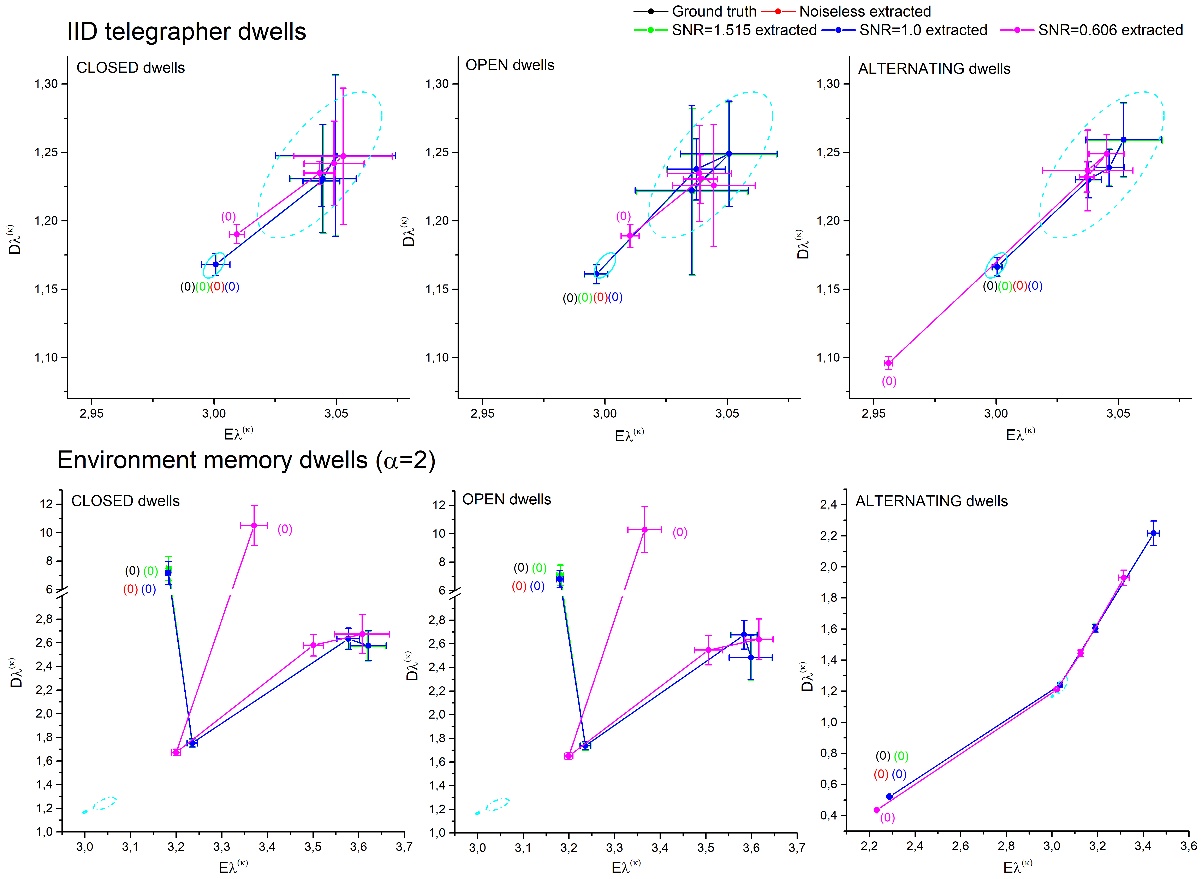}
\caption{\textbf{Effects of observational noise on extracted IID and
memory-modulated dwell sequences.}\\
Columns show closed, open and complete alternating dwell sequences. The
upper row shows IID telegrapher data; the lower row shows the
environmental-memory model with \ensuremath{\alpha}=2. Generated ground-truth dwells are
compared with dwells extracted from the noiseless binary trace and after
addition of Gaussian noise at SNRs of 1.515, 1.0 and 0.606. Noise alters
the detected event lists through additional threshold crossings, but
does not generate the progressive hierarchy displacement observed in the
memory model. Points and error bars are mean \ensuremath{\pm} SD across 12 independent
realizations per condition. Numbers identify hierarchy orders. Solid and
dashed cyan ellipses denote the 95\% IID reference regions at \ensuremath{\kappa}=0 and
\ensuremath{\kappa}=3, respectively.}
\label{fig:S8}
\end{figure}

\clearpage

\begin{figure}[!htp]
\centering
\includegraphics[width=0.92\textwidth,height=7.0in,keepaspectratio]{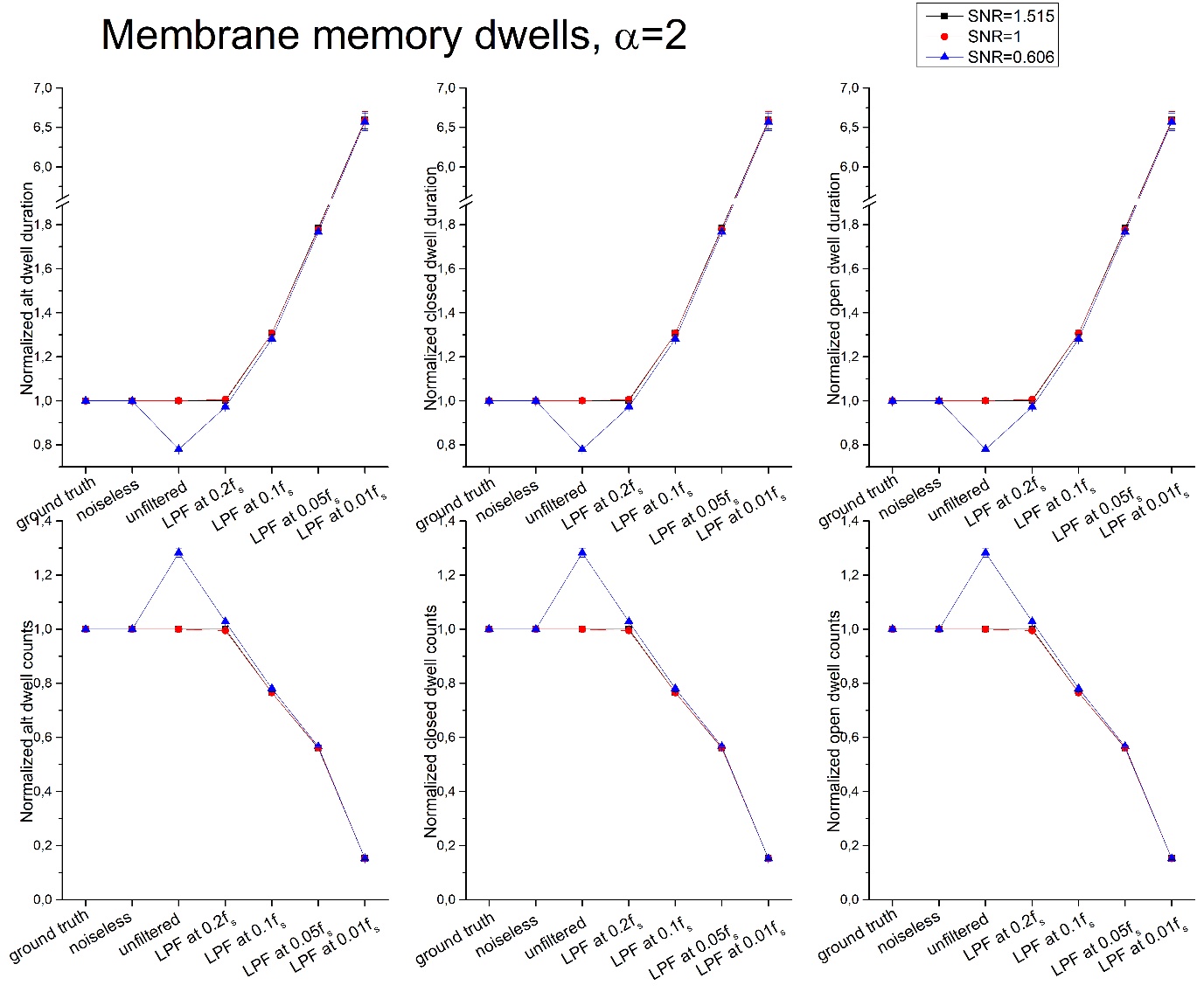}
\caption{\textbf{Low-pass filtering trades noise-induced transitions
for merged dwell events.}\\
Normalized mean dwell durations are shown in the upper row and
normalized dwell counts in the lower row for alternating, closed and
open dwells generated by the membrane-memory model with \ensuremath{\alpha}=2. Values are
normalized to the corresponding ground-truth quantities. Ground truth
and noiseless extraction are followed by unfiltered noisy traces and
traces filtered at 0.2\emph{f\textsubscript{s}},
0.1\emph{f\textsubscript{s}}, 0.05\emph{f\textsubscript{s}} and
0.01\emph{f\textsubscript{s}}. Symbols distinguish SNRs of 1.515, 1.0
and 0.606. Unfiltered noise increases detected event counts and shortens
mean durations, whereas progressively lower cutoffs reduce event counts
and increase apparent durations through event merging. Points and error
bars are mean \ensuremath{\pm} SD across 12 independent realizations.}
\label{fig:S9}
\end{figure}

\clearpage

\begin{figure}[!htp]
\centering
\includegraphics[width=0.92\textwidth,height=7.0in,keepaspectratio]{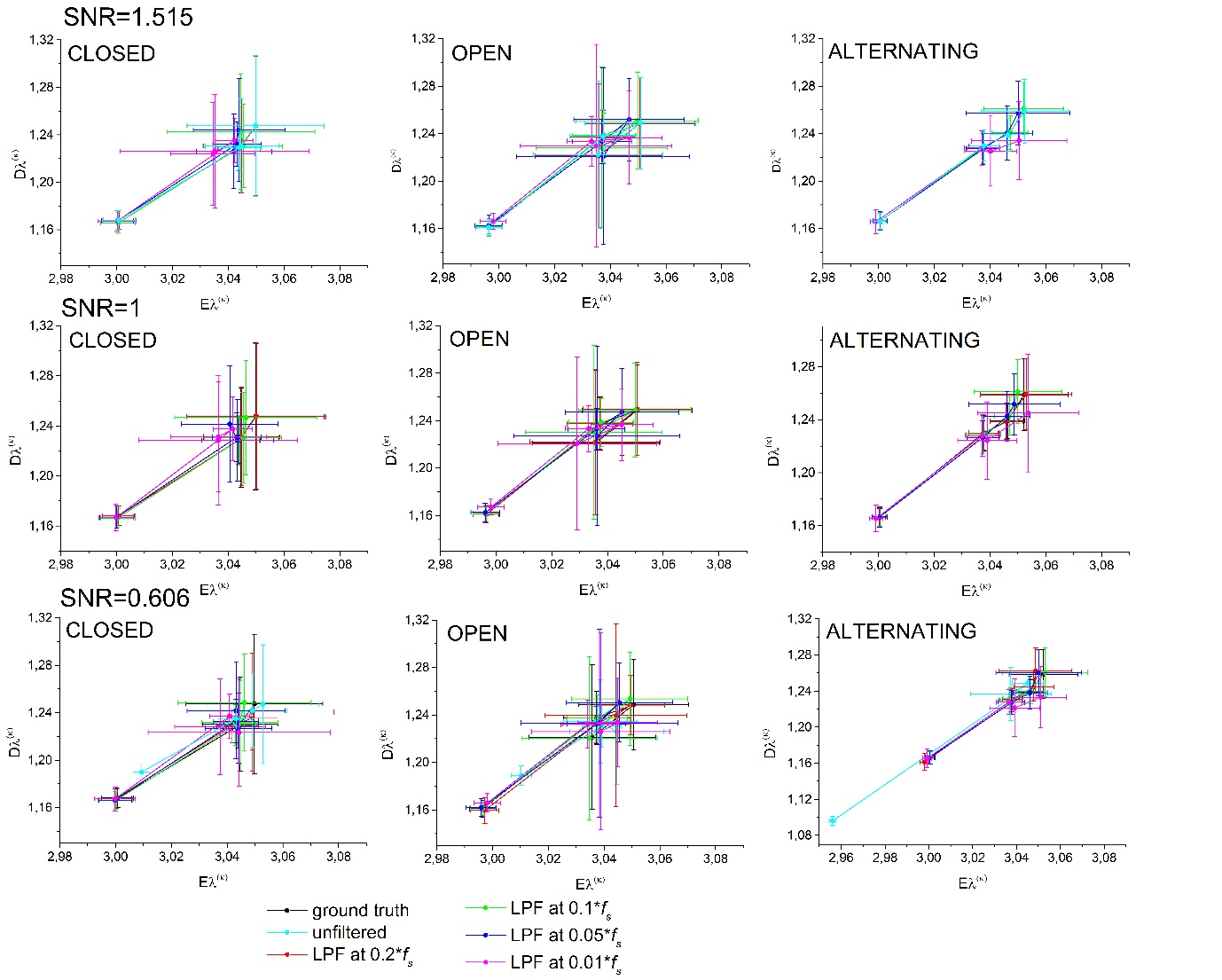}
\caption{\textbf{Hierarchy trajectories after low-pass filtering of
noisy IID dwell sequences.}\\
Rows show SNRs of 1.515, 1.0 and 0.606; columns show closed, open and
alternating dwell sequences from the IID telegrapher dwell model.
Ground-truth trajectories are compared with those extracted from
unfiltered noisy traces and after fourth-order zero-phase low-pass
filtering at 0.2\emph{f\textsubscript{s}}, 0.1\emph{f\textsubscript{s}},
0.05\emph{f\textsubscript{s}} and 0.01\emph{f\textsubscript{s}}.
Moderate filtering reduces noise-induced event-detection errors while
retaining similar hierarchy trajectories; the lowest cutoffs produce
larger displacements as short dwells are merged. Points and error bars
are mean \ensuremath{\pm} SD across 12 independent realizations per condition; where
not visible, error bars are smaller than the symbols.}
\label{fig:S10}
\end{figure}

\clearpage

\begin{figure}[!htp]
\centering
\includegraphics[width=0.92\textwidth,height=7.0in,keepaspectratio]{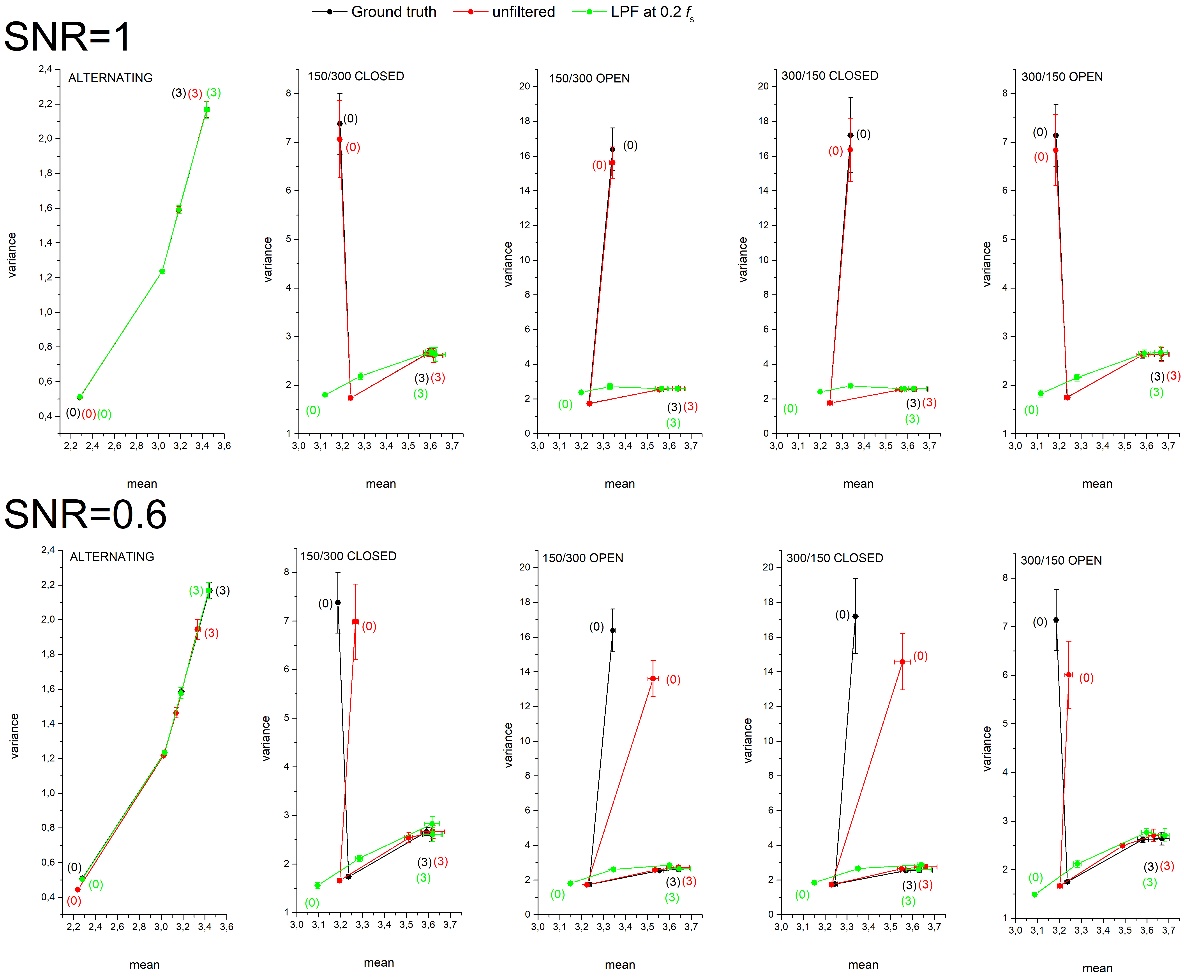}
\caption{\textbf{Effects of reciprocal open--closed dwell asymmetry
on event-derived hierarchy trajectories.}\\
The upper and lower rows show SNRs of 1.0 and 0.606, respectively, for
the strong environmental-memory model \ensuremath{\alpha}=2. The leftmost panels show the
complete alternating dwell sequence. The remaining columns show closed
and open dwells separately for the reciprocal baseline-scale conditions
150/300 and 300/150 sampling intervals. Ground-truth trajectories are
compared with trajectories extracted from unfiltered noisy traces and
after low-pass filtering at 0.2\emph{f\textsubscript{s}}. Numbers
identify \ensuremath{\kappa}=0 and \ensuremath{\kappa}=3. Dwell asymmetry has a larger effect on the
separately analysed state with the shorter characteristic duration,
whereas the alternating trajectory is comparatively stable. Points and
error bars are mean \ensuremath{\pm} SD across 12 independent realizations per
condition.}
\label{fig:S11}
\end{figure}

\clearpage

\begin{figure}[!htp]
\centering
\includegraphics[width=0.92\textwidth,height=6.7in,keepaspectratio]{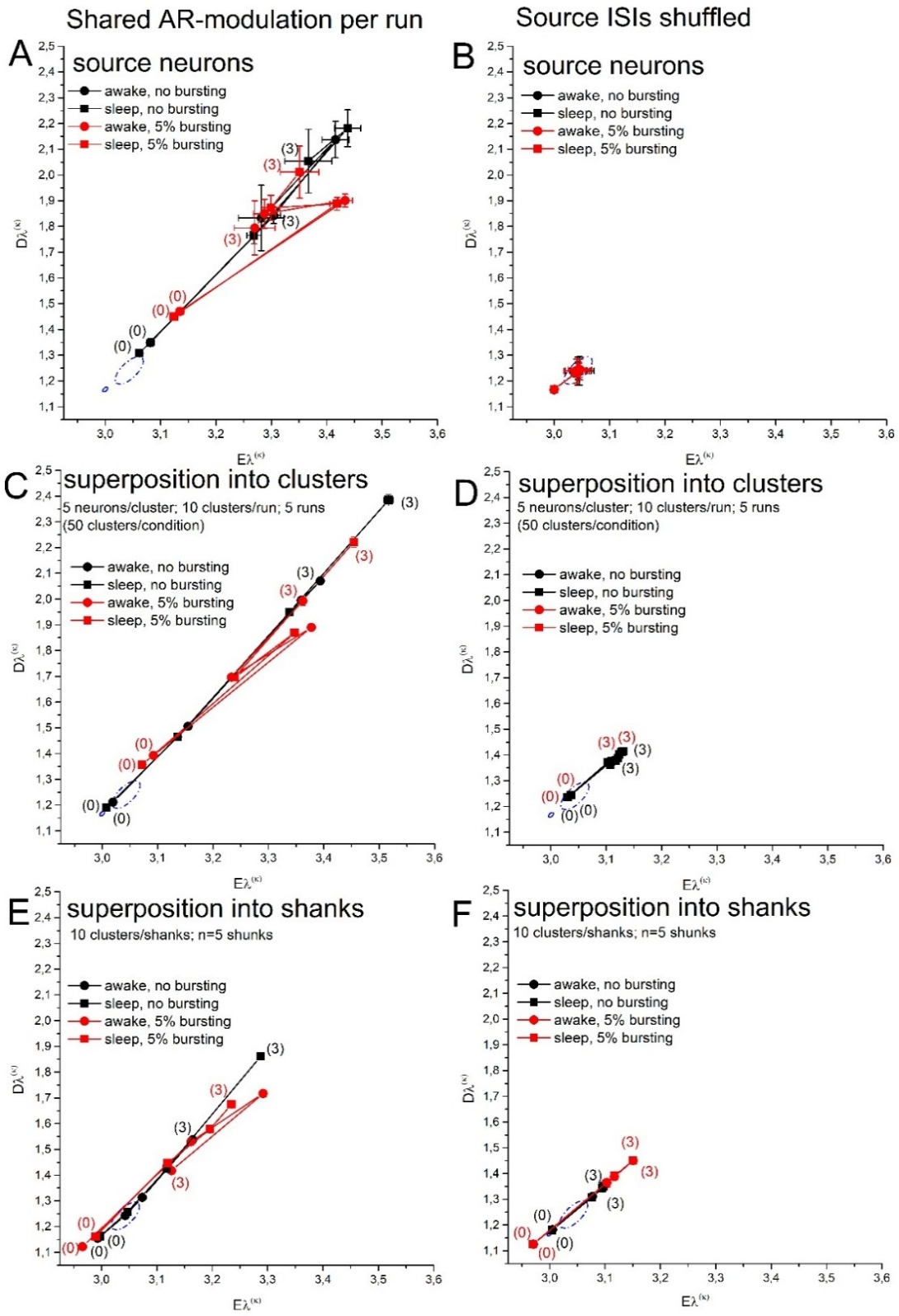}
\caption{\textbf{Hierarchy statistics of synthetic neuronal
populations with autoregressive rate modulation.}\\
\textbf{A,B,} Source-neuron ISI trajectories before and after
independent within-neuron ISI shuffling. Conditions comprise awake-like
independent modulation (c=0) and sleep-like shared modulation (c=0.75),
with or without approximately 5\% bursting. \textbf{C,D,} Corresponding
trajectories after five source neurons were superposed into each
cluster, using organized or source-shuffled neurons, respectively.
\textbf{E,F,} Corresponding trajectories after ten clusters were
superposed into each shank. Five independent population realizations
contained 50 neurons, 10 clusters and 1 shank each. Points and error
bars are mean \ensuremath{\pm} SD across 250 neurons, 50 clusters or 5 shanks per
condition. Labels identify \ensuremath{\kappa}=0 and \ensuremath{\kappa}=3. Solid and dashed blue ellipses
denote the 95\% IID reference regions at these orders.}
\label{fig:S12}
\end{figure}

\clearpage

\begin{figure}[!htp]
\centering
\includegraphics[width=0.92\textwidth,height=6.7in,keepaspectratio]{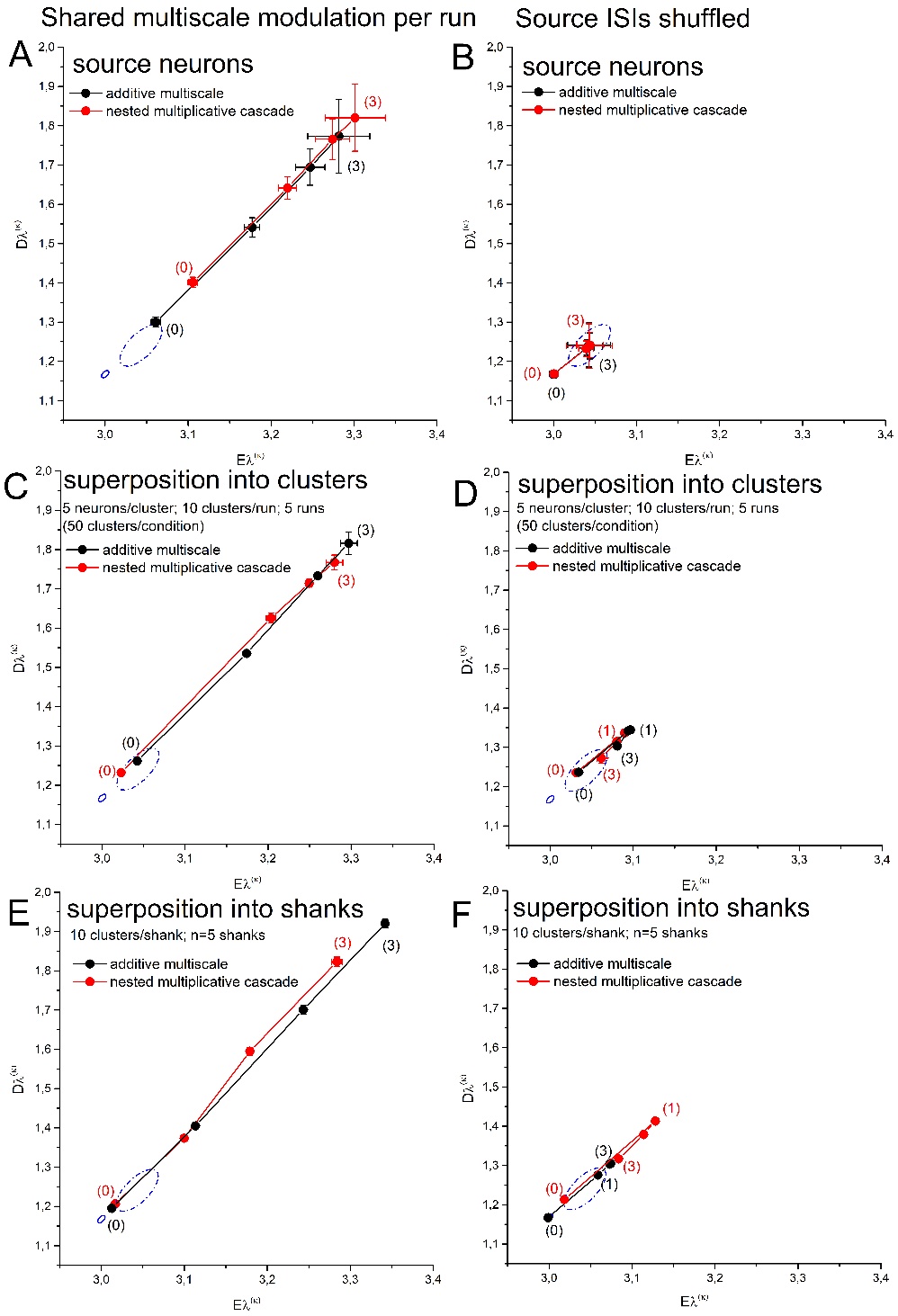}
\caption{\textbf{Hierarchy statistics of synthetic populations with shared
multiscale modulation.}\\
\textbf{A,B,} Source-neuron ISI trajectories for additive multiscale
modulation and a nested multiplicative cascade, before and after
independent within-neuron ISI shuffling. \textbf{C,D,} Corresponding
trajectories after superposition of five organized or source-shuffled
neurons into each cluster. \textbf{E,F,} Corresponding trajectories
after superposition of ten clusters into each shank. Five independent
population realizations contained 50 neurons, 10 clusters and 1 shank
each. Points and error bars are mean \ensuremath{\pm} SD across 250 neurons, 50
clusters or 5 shanks per model. Labels identify selected hierarchy
orders. Solid and dashed blue ellipses denote the 95\% IID reference
regions at \ensuremath{\kappa}=0 and \ensuremath{\kappa}=3, respectively.}
\label{fig:S13}
\end{figure}

\clearpage

\begin{figure}[!htp]
\centering
\includegraphics[width=0.92\textwidth,height=6.7in,keepaspectratio]{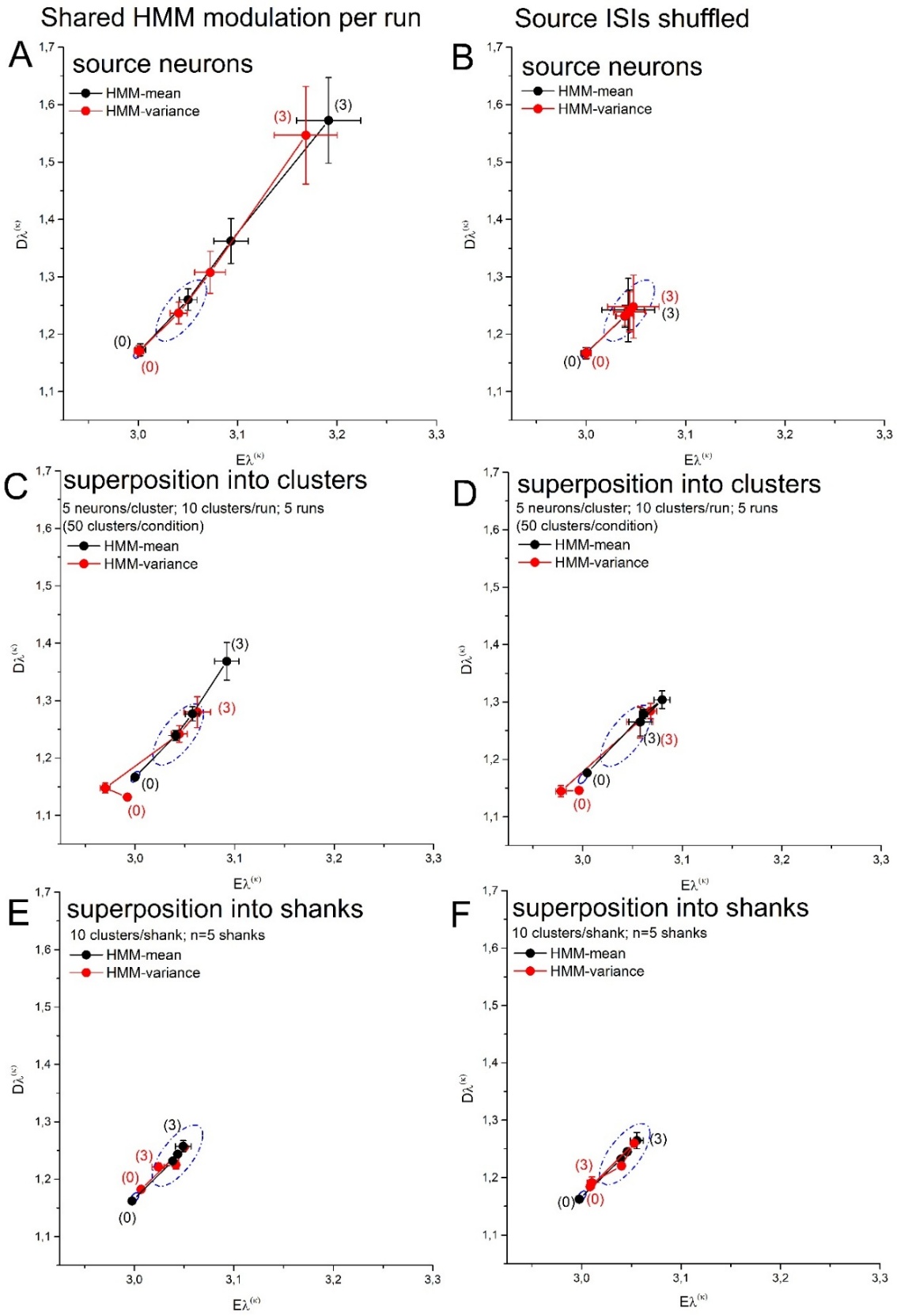}
\caption{\textbf{Hierarchy statistics of synthetic populations with
shared hidden-state modulation.}\\
\textbf{A,B,} Source-neuron ISI trajectories for mean-switching and
variance-switching hidden-Markov models, before and after independent
within-neuron ISI shuffling. \textbf{C,D,} Corresponding trajectories
after superposition of five organized or source-shuffled neurons into
each cluster. \textbf{E,F,} Corresponding trajectories after
superposition of ten clusters into each shank. Five independent
population realizations contained 50 neurons, 10 clusters and 1 shank
each. Points and error bars are mean \ensuremath{\pm} SD across 250 neurons, 50
clusters or 5 shanks per model. Labels identify \ensuremath{\kappa}=0 and \ensuremath{\kappa}=3. Solid and
dashed blue ellipses denote the 95\% IID reference regions at these
orders.}
\label{fig:S14}
\end{figure}

\clearpage

\begin{figure}[!htp]
\centering
\includegraphics[width=0.92\textwidth,height=6.7in,keepaspectratio]{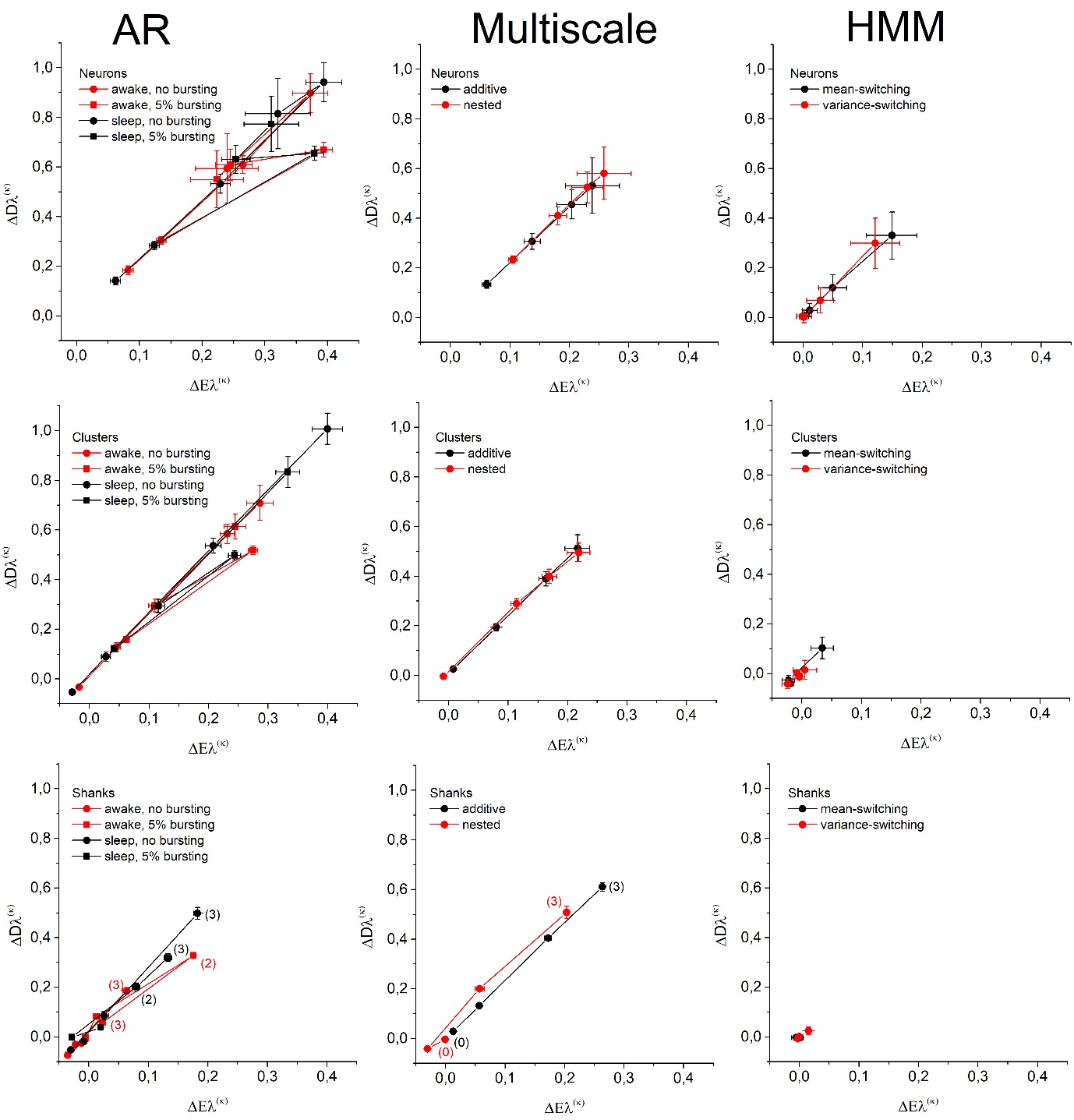}
\caption{\textbf{Organized-minus-null residual trajectories across
neuronal aggregation levels.}\\
Columns show autoregressive, multiscale and hidden-Markov population
models; rows show source neurons, five-neuron clusters and ten-cluster
shanks. Residual coordinates were calculated at each hierarchy order as
\(\Delta E\lambda^{(\kappa)} = E\lambda_{org}^{(\kappa)} - E\lambda_{null}^{(\kappa)}\)
and
\(\Delta D\lambda^{(\kappa)} = D\lambda_{org}^{(\kappa)} - D\lambda_{null}^{(\kappa)}\).
The primary null independently shuffled every source neuron's ISIs
before repeating the complete superposition procedure. In the AR column,
colour distinguishes bursting and marker shape distinguishes awake-like
(c=0) and sleep-like (c=0.75) conditions (see Methods). The multiscale
column compares additive and nested modulation; the HMM column compares
mean- and variance-switching models. Lines connect \ensuremath{\kappa}=0--3; order labels
are shown only where needed to identify shank-level excursions. Points
and error bars are paired mean \ensuremath{\pm} sample SD across 250 neurons, 50
clusters or 5 shanks per model and order.}
\label{fig:S15}
\end{figure}

\clearpage

\begin{figure}[!htp]
\centering
\includegraphics[width=0.92\textwidth,height=7.0in,keepaspectratio]{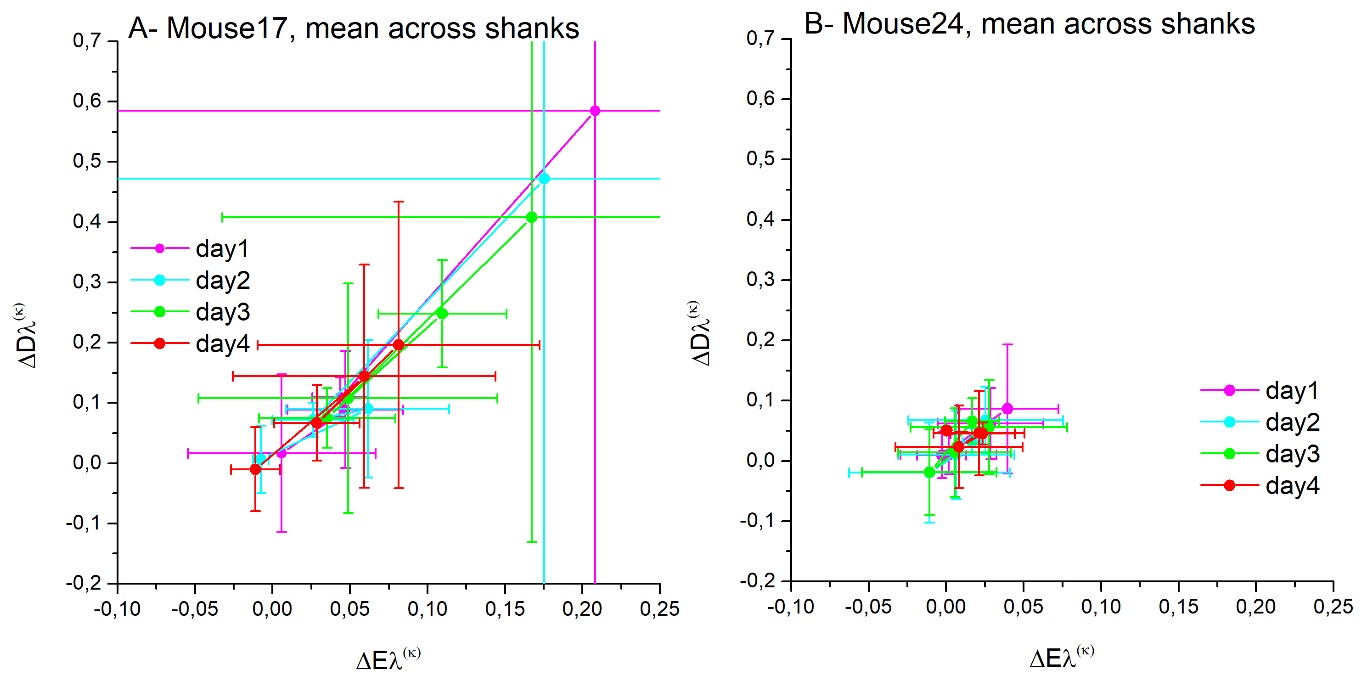}
\caption{\textbf{Across-day residual hierarchy trajectories in Mouse
17 and Mouse 24.}\\
\textbf{A,B,} Mean real-minus-null residual trajectories for Mouse 17
and Mouse 24, respectively, during slow-wave sleep. For every day \ensuremath{\times}
shank record, each constituent cluster's within-interval ISIs were
independently shuffled five times before reconstruction and
superposition. The centroid of these five matched nulls was subtracted
from the recorded shank coordinate at each hierarchy order. Colours
identify chronologically ordered recording days, and lines connect k:
(0) \ensuremath{\rightarrow} (1) \ensuremath{\rightarrow} (2) \ensuremath{\rightarrow} (3). Points and error bars show mean \ensuremath{\pm} SD across
eligible shanks within each recording day. Only paired recorded and null
trajectories satisfying the order-specific inclusion criteria were used.}
\label{fig:S16}
\end{figure}

\clearpage
\end{document}